\documentclass[journal]{IEEEtran}
\usepackage{ifpdf}
\usepackage{cite}
\usepackage{amsmath}
\usepackage{array}
\usepackage[caption=false]{subfig}
\usepackage{cite}
\usepackage{graphicx}
\usepackage{multirow}
\usepackage{amsmath}
\usepackage{epsfig}
\usepackage[center]{caption}
\usepackage{epstopdf}
\usepackage{lscape}
\usepackage{algorithm}
\usepackage{url}
\usepackage{algorithmicx}
\usepackage{algpseudocode}
\usepackage{blkarray}
\usepackage[table,xcdraw]{xcolor}
\usepackage{float}
\usepackage{amsfonts}
\usepackage{mathrsfs}
\usepackage[utf8x]{inputenc}

\begin{document}

\title{Comparing H.265/HEVC and VP9: Impact of High Frame Rates on the Perceptual Quality of Compressed Videos}

\author{Tariq~Rahim,~\IEEEmembership{Member,~IEEE,}
          and Muhammad~Arslan Usman,~\IEEEmembership{Member,~IEEE,}
       \\~and Soo~Young Shin,~\IEEEmembership{Senior Member~,~IEEE}
\thanks{T. Rahim and S. Y. Shin are with the WENS Laboratory, Department of IT Convergence
	Engineering, Kumoh National Institute of Technology, Gumi 39177, South
	Korea (e-mail: tariqrahim@ieee.org; wdragon@kumoh.ac.kr).}
\thanks{Muhammad Arslan Usman is with working jointly with Kingston University London and Pangea Connected London under the umbrella of Knowledge Transfer Partnerships (KTP), United Kingdom.}
\thanks{Manuscript received}}

%
%

\markboth{Pre-print submitted to arxiv.org}%
{Rahim \MakeLowercase{\textit{et al.}}: Bare Demo of IEEEtran.cls for IEEE Journals}

\maketitle

\begin{abstract}
High frame rates have been known to enhance the perceived visual quality of specific video content. However, the lack of investigation of high frame rates has restricted the expansion of this research field—particularly in the context of full-high-definition (FHD) and 4K ultra-high-definition video formats. This study involves a subjective and objective quality assessment of compressed FHD videos. First, we compress the FHD videos by employing high-efficiency video coding, and VP9 at five quantization parameter levels for multiple frame rates, i.e. 15fps, 30fps, and 60fps. The FHD videos are obtained from a high frame-rate video database BVI-HFR, spanning various scenes, colors, and motions, and are shown to be representative of the BBC broadcast content. Second, a detailed subjective quality assessment of compressed videos for both encoders and individual frame rates is conducted, resulting in subjective measurements in the form of the differential mean opinion score reflecting the quality of experience. In particular, the aim is to investigate the impact of compression on the perceptual quality of compressed FHD videos and compare the performance of both encoders for each frame rate. Finally, 11 state-of-the-art objective quality assessment metrics are benchmarked using the subjective measurements, to investigate the correlation as a statistical evaluation between the two models in terms of correlation coefficients. A recommendation for enhancing the quality estimation of full-reference (FR) video quality measurements (VQMs) is presented after the extensive investigation.  
\end{abstract}


%
\IEEEpeerreviewmaketitle

\section{Introduction}
\IEEEPARstart{W}{ith} the progress in technology for capturing, storing, transmitting, and displaying video content, high-quality video services have  recently become prevalent. Nowadays, video content at high-definition (HD) resolutions are provided by most broadcasting companies and online video web sites. Following the success of HD video services, the 4K ultra-high-definition (UHD) resolution format with $3840\times2160$ pixels is regarded as the future standard in video applications \cite{cheon2017subjective}. Recently, there has been an increased focus on the implementation of high-spatial-resoultion (4K/8K), high-dynamic-range (HDR), and immersive multi-view formats. Still, the progress related to high-frame-rate formats has been relatively slow, as evident from the frame rates for entertainment videos, such as cinematic films and TV programs, whose resolution seldom surpasses 60fps \cite{mackin2018study}.
\par 
Ideally, full-high-definition (FHD) content is expected to provide viewers with improved visual experience through a wide field of view both horizontally and vertically, with suitable screen sizes. FHD with $1920\times1080$ pixels has two times the spatial resolution than HD Ready, and thus, can deliver a larger amount of visual information to viewers. This increase in resolution is the initial stage of an immersive and naturalistic visual experience \cite{usman2018exploiting}. 
\par 
Recently, high frame rates (HFRs) have stimulated interest in communities such as broadcast, film (Avatar, Billy Lynn’s Long Halftime Walk), online streaming, virtual reality, and gaming. For the ultra-high-definition video standard (Rec. 2020), up to 120fps have been specified \cite{bt20202012parameter}. However, the need for higher-resolution and HFR videos is growing because of the availability of 4K and 8K UHD contents and larger display screens. Although frame interpolation and different post-processing methods can alleviate the artifacts found in low frame rates, satisfactory results have not been obtained. Because of many dynamics in a sequence, human viewers do not require a fixed HFR for a full video. For instance, low dynamics can be achieved for a video with lower coding frame rate \cite{huang2016perceptual}.    
\par 
Nevertheless, the FHD format produces a new challenge; that is, managing the increased amount of data in FHD video services needs more storage capacity and bandwidth. To address this problem, video compression with essential measure is necessary. The High-Efficiency Video Coding (HEVC) \cite{sullivan2012overview} is the latest standard for video compression, developed by the Joint Collaborative Team on Video Coding, which was founded by both the ITU-T Video Coding Experts Group  and the ISO/IEC Moving Pictures Expert Group in 2010. In 2013, the final HEVC specification was approved by ITU-T as Recommendation H.265 and by ISO/IEC as MPEG-H, Part 2. Part of the WebM project, VP9 \cite{bankoski2013towards} is another open-source compression method introduced recently. For any encoder, the compression level and variation in the frame rate have a direct relation with the quality of video content. Hence, it is necessary to investigate the impact of compression level on the perceptual quality of users for different frame rates using different encoders.     
\par 
Before the exploitation of HFR for future video formats, further research is required to identify the contribution made by the different frame rates for the entire video pipeline, i.e., from obtaining the video through compression till transmission for visual perception. Various attempts have been made to determine the relation between frame rate and perceived quality, such as examination of motion blurring perception \cite{sugawara2009p} at a frame rate of 30Hz. Similarly, a subjective test was conducted by utilizing QCIF and CIF for frame rates below 30fps, to investigate the impact of frame rate and quantization parameter on the perceptual quality of video \cite{ou2010perceptual}. The exploitation of frame rates above 30fps are quite rare, and some studies have been reported \cite{mackin2015study,mackin2018study}, and \cite{nasiri2017perceptual} to be concentrating on either high-resolution videos, such as 4K, or a gaming environment.   
\par 
Meanwhile, very few HFR databases have been publicly released thus limiting the research, which also implies that robust inferences regarding HFR are difficult. The available databases are mostly based on either a single frame rate or low frame rates. To cope them, a publicly available HFR video database Bristol Vision Institute High Frame Rate (BVI-HFR) \cite{mackin2015study} containing 22 diverse uncompressed FHD video sequences with a resolution of $1920\times1080$, each with 10 sec duration, and source videos of 120fps is utilized.
\par
In the context of this paper, we utilize the BVI-HFR database to investigate the impact of compression on perceptual quality by encoding the video contents with H.265/HEVC and VP9 encoders using five different QP levels. A detailed subjective quality assessment for the compressed videos for both encoders and individual frame rates is conducted, resulting in subjective measurements in the form of DMOS reflecting the quality of experience. In particular, the aim is to investigate and compare the impact of compression on the perceptual quality of compressed FHD videos. Then 11 state-of-the-art objective quality assessment metrics are benchmarked using the subjective measurements, to investigate the correlation as a statistical evaluation between the two models in terms of correlation coefficients. This study aims to investigate the performance of the opted encoders with different frame rates under different QP levels. The BVI-HFR video database has a native video of 120 fps, which is temporally down-sampled by averaging to 60fps, 30fps, and 15fps \cite{mackin2017investigating}. We use the video contents of frame rates 15fps, 30fps, and 60fps for our investigation.   
\par 
The major contributions of this study are outlined as follows:
\begin{itemize}
\item First, compression of FHD video contents using H.265/HEVC and VP9 at five QP levels 27, 31, 35, 39, and 43 at frame rates of 15fps, 30fps, and 60fps is conducted. The compression of the FHD video content is separately performed for each frame rate for both encoders. 
\item Second, a detailed subjective quality assessment for the compressed video contents at 15fps, 30fps, and 60fps is conducted to generate DMOS values reflecting the perceptual quality of the users.  
\item Employing 11 state-of-the-art FR objective-VQA metrics, we attempt to quantify the relation between subjective measurements i.e. DMOS, and FR-VQA metrics. The aim is to investigate the impact of frame rate variations on the perceptual quality and performance comparison of the opted encoders at different QP levels. After conducting statistical evaluation to validate both models in terms of correlation coefficients (cc), FR-VQA metrics for both H.265/HEVC and VP9 is recommended for compressed FHD contents.   
\item Finally, a recommendation for enhancing the quality estimation of full-reference (FR) video quality measurements (VQMs) is presented after the extensive investigation.  
\end{itemize}

\section{RELATED WORK}
\subsection{Advantages of Increased Frame Rate}
Previous research has shown that there are several  distinct advantages associated with increased frame rates: enhanced depth perception for both non-expert \cite{wilcox2015evidence} and expert \cite{allison2017expert} viewers; improved realism; more constant motion; decrease in perceptible motion blurring \cite{emoto2014high}; diminishing of temporal aliasing artifacts visibility \cite{watson2013high} for up to 240fps, perceptual quality improvement \cite{emoto2014high}; enhancement in spatial and speed discrimination \cite{kime2016psychophysical}; higher realistic picture quality \cite{kuroki2014effects}; and reduction in stress levels for the viewer (implied by a low blinking frequency \cite{haak2009detecting}). HFR also improves the capability of capturing slow-motion playback videos \cite{debattista2018frame}. An experimental setup is shown in \cite{mackin2018study} for fully eradicating artifacts of temporal aliasing in some scenarios at frame rates are close to 900fps.    
\par 
However, despite these advantages, HFR contents may barely be desirable in representing a ``hyper-realistic" scene (e.g., sports programming), as lower frame rates may cause a conflict with the ``cinematic appearance". Content and director providers currently have limited compliance in this matter (as in legacy formats, frame rates have for several years remained static),  and consequently, the selection of frame rates - enabled by the application of temporal down-sampling techniques can be regarded as an artistic option \cite{mackin2018study}.   
\subsection{Video Databases - High Frame Rates (HFR)}
Very few HFRs of FHD video databases are publicly available \cite{mackin2018study}. Previous studies used either single frame rates or comparatively low frame rates, i.e, (24fps \cite{barman2017h}, 30fps \cite{ou2008modeling, huang2016perceptual, ou2014q, nightingale2014impact, ma2011modeling, cheon2017subjective}. In contrast, few studies have focused on frame rates above 50fps \cite{nightingale2014impact}, and 60fps \cite{huang2016perceptual}. 
\subsection{Video Compression and Configuration}
Frame-rate-related video quality and compression for different QP values have been studied for almost two decades. These efforts can be roughly categorized into two main classes based on their goals. The first class is concerned with various viewing positions and artifacts perceived by viewing the video aired at different frame rates. The second one concentrates on efficient video compression techniques for decreasing the coding bit rates with little quality degradation \cite{huang2016perceptual}. 
\begin{figure*}[t]
	\centering
	\includegraphics[width=7.0in,height=7.0in,keepaspectratio]{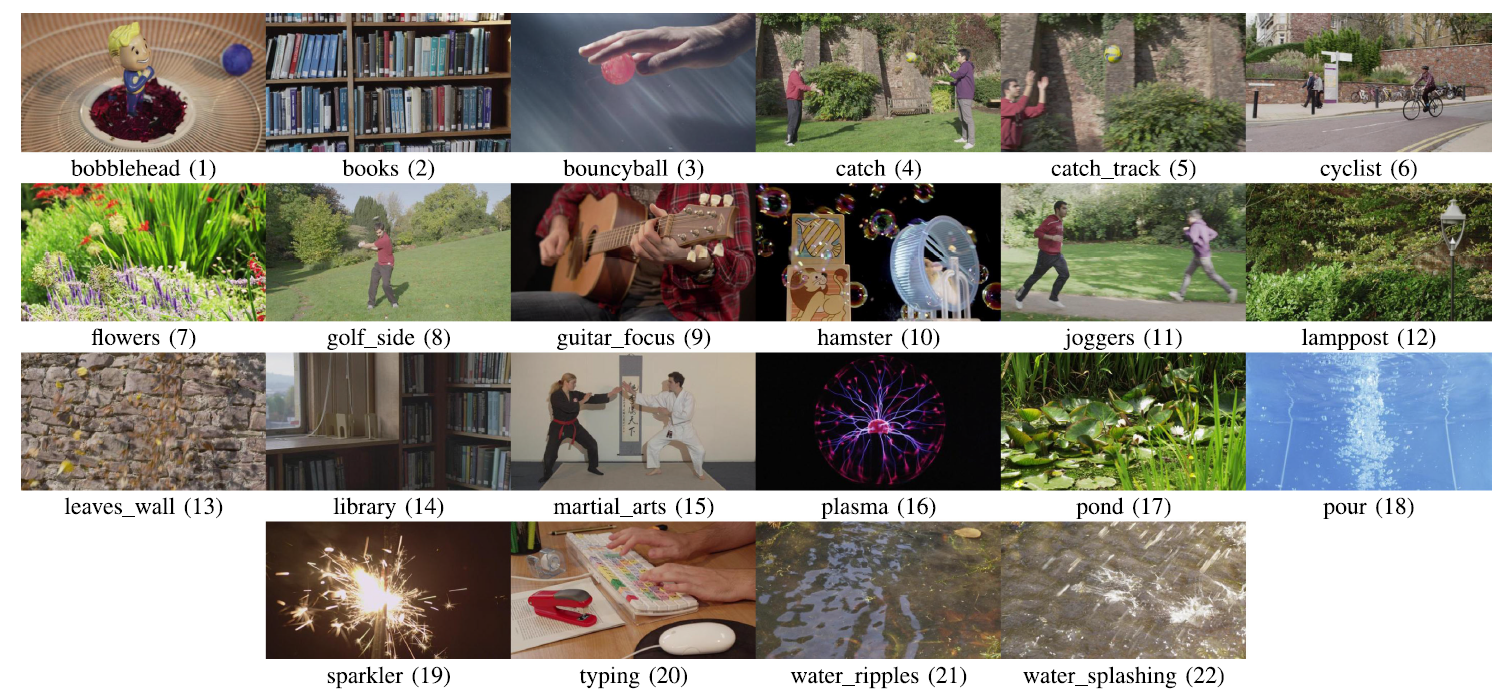}
	\caption {A sample frame from each of the 22 video sequences in the BVI-HFR video database, along with the names and associated indices.}
	\label{Fig4}
\end{figure*}
\par 
To investigate the impact of frame rate and QP on perceptual quality of a video, the product of spatial quality factor (SQF) and a temporal correction factor (TCF) is utilized \cite{ou2010perceptual}. SQF estimates the decoded frames quality while TCF decreases the quality defined by the initial factor according to the original frames rate. A high correlation is achieved between subjective assessment and the proposed content-dependent metric \cite{ou2010perceptual}. Subjective analysis is conducted to evaluate the effect of frame rate and H.264 compression on the  perceived video quality. A direct relation between the frame rate and video quality is shown; yet the dependencies on QP level, spatial resolution, and video content statistics are important \cite{nasiri2017perceptual}. A rate-distortion optimization that adaptively determines QPs for a group of neighboring frames, mostly implemented in H.265/HEVC for decreasing the coding distortion, resulted in sufficient minimization of the BD-rate and quality fluctuation \cite{he2017adaptive}.                   
\par 
Besides video compression, there has also been considerable recent progress with respect to different codec comparisons. An objective analysis for evaluating the performance comparison of H.264/MPEG-AVC, H.265/MPEG-HEVC, and VP9 video encoders utilizing gaming videos for live streaming applications was conducted in \cite{barman2017h}. A subjective and objective assessment for UHD and TV broadcast situations \cite{vrevrabek2014comparison} was conducted to investigate the coding efficiency of HEVC, AVC, and VP9, indicating that HEVC outperforms the other encoders. For UHD, FHD, and HD videos \cite{bienik2016performance} the coding efficiency and quality at the end-user were examined for codecs such as H.264, H.265, VP8, and VP9. Similarly, a detailed subjective and objective analysis \cite{vrevrabek2015quality} was conducted for real-time applications using HEVC and VP9 encoders showing better performance of HEVC in terms of compression.          
\section{Video Database and Encoding Configuration}
A HFR, 120fps from the BVI-HFR video database \cite{mackin2015study} was used for comparing the performance of H.265/HEVC and VP9 encoders to investigate the impact of HFR on the perceptual quality of compressed videos. 
\subsection{Video Database}
The BVI-HFR video database \cite{mackin2015study} comprises 22 unique uncompressed video sequences at FHD resolution ($1920\times1080$), with duration of 10 sec, and frame rate of 120fps. Each video sequence was further been temporally down-sampled by averaging frames to 60fps, 30fps, and 15fps - resulting in a total of 88 sequences. Fig. 1 depicts the sample frame of the database with associated name and index open to download from the link: https://vilab.blogs.ilrt.org/?p=1563.
\subsection{Content Description}
The encoding complexity of a video sequence and the compression difficulty are based on the complexity of the content, which is defined by employing  Spatial Information (SI) and Temporal Information (TI). SI measures the amount of edge energy that can be used to measure the spatial details, while TI predicts the magnitude of temporal changes. Fig. 2 presents the spatial and temporal contents of the videos for the BVI-HFR database. All 22 source sequences at a frame rate of 120fps are used to measure the SI and TI descriptions for the BVI-HFR database. Based on the SI and TI results, five video sequences from the database were selected for the assessment \cite{mackin2018study}. From the database, the temporally down-sampled frame rates of 60fps, 30fps, and 15fps were opted for investigating the impact of compression by encoding the videos with H.265/HEVC and VP9 encoders under different QP values.  
\begin{table*}[]
	\centering
	\caption{Encoders settings and configuration for different QP levels}
	\renewcommand{\arraystretch}{1.15}
	\begin{tabular}{|c|c|c|}
		\hline
		\textbf{Codec}      & \textbf{Version}        & \textbf{Parameters}                                                                                                                                                                                                                                                                                                                                                                                                    \\ \hline
		\textbf{H.265/HEVC} & \text{libx265 v2.7.0} & \text{\begin{tabular}[c]{@{}c@{}} \\ ffmpeg -i (INPUT) -c:v  libx265 -x265-params  pass=1  -strict experimental -b:v 8000k\\  -minrate 800k -maxrate 8000k  -pix\_fmt yuv 422p medium -an -f mp4 /dev/null \\ \\ ffmpeg -i (INPUT) -c:v  libx265  -x265-params passs=2  qp=(27 to 43) -c:a aac  -strict experimental -b:v 8000k\\  -minrate 800k -maxrate 8000k  -pix\_fmt yuv 422p medium  -an output.mp4\end{tabular}} \\ \hline
		\textbf{VP9/WebM}   & \text{libvpx  v1.7.0} & \text{\begin{tabular}[c]{@{}c@{}} \\ ffmpeg -y -i (INPUT) -c:v libvpx-vp9 -b:v 8000k -pass 1 -c:a opus -b:a 64k -f webm /dev/null\\ ffmpeg -i (INPUT\_Pass1) -c:v libvpx-vp9 -b:v 8000k -pass 2 -c:a opus -b:a -qmin 24 -qmax 26 (\%For 27) \\ -f webm output.webm\end{tabular}}                                                                                                                                         \\ \hline
	\end{tabular}
\end{table*}

\subsection{Video Compression and Encoder Settings}
For our analysis, temporally down-sampled frame rates of 60fps, 30fps, and 15fps were selected from the BVI-HFR database, resulting in a total 66 video sequences. All video sequences were FHD with a resolution of $1920\times1080$ in the eight-bit YUV 4:2:0 format. The resulting 66 video sequences for each frame rate were then encoded using HEVC and VP9 at five QPs: 27, 31, 35, 39, and 43. For example, the video sequences at frame rate 60fps, five videos were selected based on the SI and TI plotting as shown in Fig. 2 and were encoded by HEVC at five QP levels, resulting in 25 encoded sequences. The same process for frame rate 30fps and 60fps resulting in a total of 75 encoded sequences. The same encoding process was used for VP9 resulting in a total 75 encoded video sequences. Therefore, 150 encoded video sequences are achieved employing both encoders.
\begin{figure}[H]
	\centering
	\includegraphics[width=3.45in,height=7.00in,keepaspectratio]{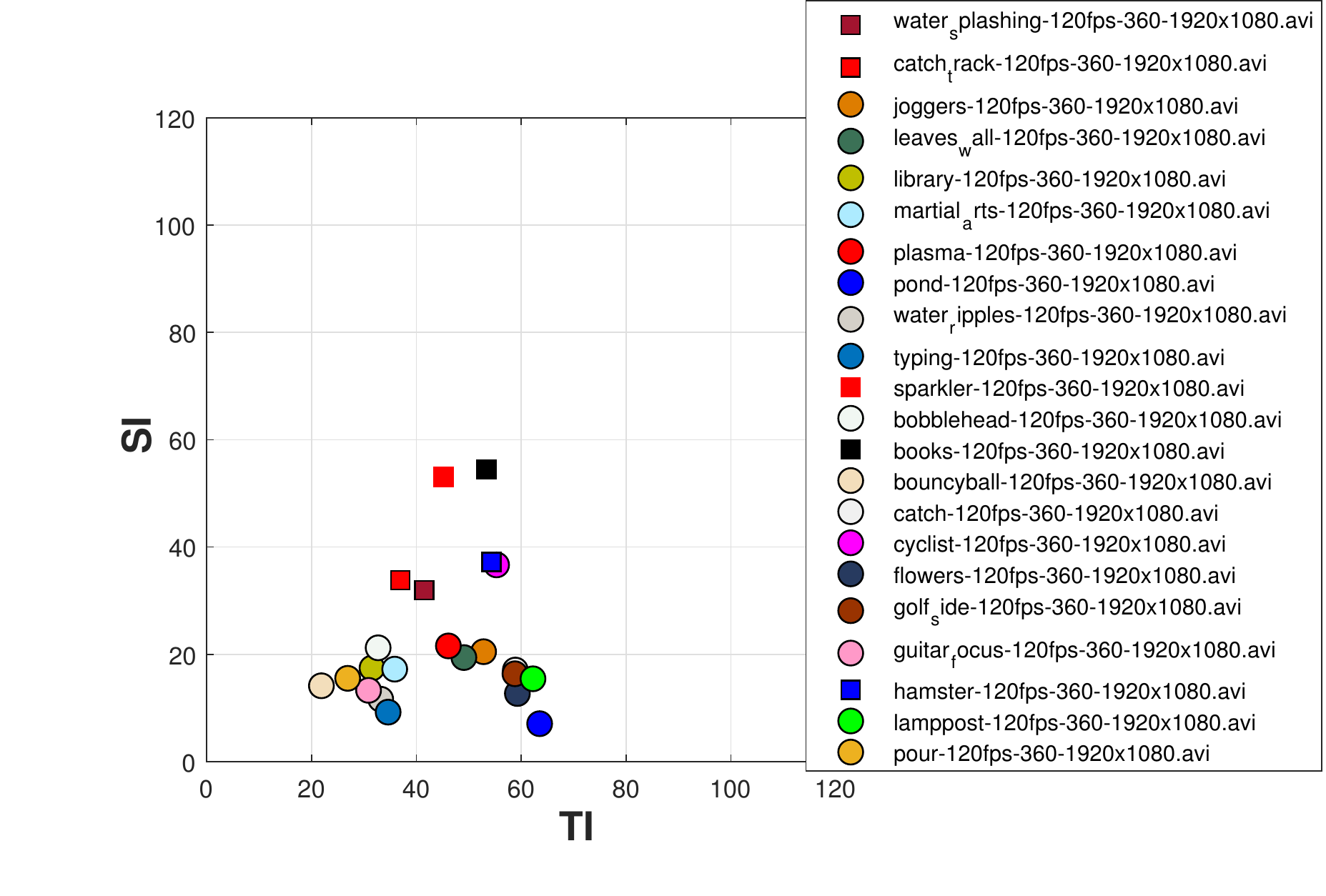}
	\caption{Spatial-Temporal plot for BVI-HFR database}
	\label{Fig4}
\end{figure}
\par
In our study, the FFmpeg open-source libraries, libx265 and libvpx-vp9, were used as the encoder wrapper for H.265/HEVC and VP9 codec, respectively. The details of the encoder settings are listed in Table. I. Here, for balancing both encoder configurations, modifications such as, the selection of a preset instead of \text{\textit{ultrafast}} and \text{\textit{veryfast}}, the \text{\textit{medium}}  preset were performed for the same encoding efficiency and quality but at the cost of compression efficiency.  
\section{Subjective Evaluation}
This section details the experimental analysis conducted for the quantification to find the relation between QP and perceptual quality.
\subsection{Video Contents}
This section explains the techniques and materials used for conducting the subjective tests for the specific investigation. The participants for the particular research were selected based on ITU recommendations, such as ITU-R BT.500-13 \cite{series2012methodology} and ITU-T P.910 \cite{itu1999subjective}. Five video sequences were selected after analyzing the information generated from the SI and TI plotting shown in Fig. 2. From the BVI-HFR database temporally down-sampled frame rates of 60fps, 30fps, and 15fps, each containing 22 video sequences were selected and encoded via H.265/HEVC and VP9 at the five QP levels discussed in section III. 
\par 
In the rest of the paper, for the ease of usage the terms will be defined as:
\begin{itemize}
	\item \text{\textit{Source sequence (SRC)}}: An original or unimpaired/uncompressed video sequence.
	\item \text{\textit{Encoded video sequence (EVS)}}: An encoded/compressed or impaired video sequence.
	\item \text{\textit{Clip}}: Can be any video sequence i.e., either SRC or EVS.
\end{itemize}
\begin{table}[H]
	\renewcommand{\arraystretch}{1.28}
	\caption{Opinion score rating scale}
	\resizebox{9.00cm}{!}{
		\begin{tabular}{|c|c|c|c|}
			\hline
			\multicolumn{2}{|c|}{\textbf{Category Rating}}                                                            & \textbf{\begin{tabular}[c]{@{}c@{}}Opinion \\ Score\end{tabular}} & \textbf{\begin{tabular}[c]{@{}c@{}}Normalized \\ Scores\end{tabular}} \\ \hline
			\textbf{Visual Quality} & \textbf{Error Perceptibility}                                                   & \multicolumn{2}{l|}{}                                                                                                                     \\ \hline
			\text{Excellent}      & \text{Imperceptible}                                                          & \text{5}                                                        & \text{80-100}                                                       \\ \hline
			\text{Good}           & \text{\begin{tabular}[c]{@{}c@{}}Perceptible but not\\ annoying\end{tabular}} & \text{4}                                                        & \text{60-80}                                                        \\ \hline
			\text{Fair}           & \text{Slightly annoying}                                                      & \text{3}                                                        & \text{40-60}                                                        \\ \hline
			\text{Poor}           & \text{Annoying}                                                               & \text{2}                                                        & \text{20-40}                                                        \\ \hline
			\text{Bad}            & \text{Very annoying}                                                          & \text{1}                                                        & \text{0-20}                                                         \\ \hline
		\end{tabular}
	}
\end{table}
\subsection{Experimental Setup and Approach}
For the subjective assessment, a lab was specifically designed that contained only materials relevant for the tests based on ITU-T P.910 recommendations \cite{itu1999subjective}. A calibrated Newsync X24C LCD monitor with a spatial resolution of $1920\times1080$ (24inches), peak luminance of 300cd / $m^2$, refresh rate of 144 Hz, and static contrast ratio of 1,000:1 was employed for a specific experiment. A wireless mouse was accompanied  with the LCD monitor. A high-performance system equipped with the subjective video quality assessment (VQA) software provided by Moscow State University (MSU) \cite{Anu:2013}, was utilized in the test. This software is freely accessible for research and educational purposes. The viewing distance between the participant and the LCD monitor was maintained at 76cm according to ITU-T P.910 recommendations \cite{itu1999subjective}. 
\par 
Before the experiment, a training session was conducted for the adoption of each participant with the testing process. The testing process comprised sessions such as the methodology of testing, during answering any type of concerns or queries. The participants were shown two video sequences of the same specifications but not from the BVI-HFR database, and subjective scores were recorded via the below discussed approach. A total of 18 participants with an average of ($\pm$ $\sigma$)  24.6 $\pm$ 4.3 years at the Wireless Emerging Network systems (WENS) lab, were selected for the subjective tests conducted for this analysis. The selected participants for the subjective assessment were of mixed gender, including nine doctoral and nine master's students. One complete session did not last longer than 30 min, and involved 66 video sequences using the Double Stimulus Continuous Quality Scale Type II (DSCQS-II) for scoring the opinion. 
\subsection{Opinion Scores Analysis}
For evaluating the clip quality using the DSCQS Type-II method, a continuous rating value in terms of opinion scores of 0-100 (from imperceptible to very annoying) was recorded, as shown in Table. II. The DSCQS Type-II method \cite{series2012methodology} recorded the opinion scores on a five point vertical scale recommended by  ITU-R BT 500-12 for each participant, each frame rate i.e., 15fps, 30fps, and 60fps, and each encoder i.e H.265/HEVC and VP9. Each participant was shown two clips simultaneously: an SRC and an EVS clip; however, the participants were unaware of the clip type. After the participants viewed the clips, they were asked to rate the score separately as the opinion score (OS).  
\subsection{Scoring Method}
As explained earlier, in this study, we have applied the DSCQS Type-II method for subjective evaluation. The OS recorded on the five-point rating scale were transformed to a normalized scale that ranged between 0 and 100. DMOS are typically used for investigation, and calculated from mean opinion score (MOS) as follows:   
\begin{equation}
DMOS_ith = {MOS}_{ith}^{SRC} - {MOS}_{ith}^{EVS}
\end{equation}
\par
where ${MOS}_{ith}^{SRC}$ and  ${MOS}_{ith}^{EVS}$ are the calculated MOSs of the source and encoded video sequences, respectively. MOSs for each frame rate and clip were calculated as OS by each participant, and can be expressed in the generalized form as follows: 
\begin{equation}
MOS_ith = \frac{1}{N} \sum_{i=1}^{N}  {OS}_{ith}^{vth}
\end{equation}
\par 
where $N$ is the total number of participants involved in the subjective testing, and ${OS}_{ith}$ is the recorded normalized opinion score for all \textit{i}= 1,2,3....\textit{N} test subjects for the ${vth}$ clip. Fig. 3 and Fig. 4 details the subjective test conducted by 18 participants for the five selected clips at three frame rates, i.e., 15fps, 30fps, and 60fps, encoded with H.265/HEVC and VP9 at five QP (27,31,35,39, and 43) levels in the form of DMOS vs. QP. 
\section{Full-Reference Video Quality Assessment(VQA) metrics}
This section briefly explains the full-reference (FR)-objective VQA metrics used in our analysis for different frame rates.
\subsection{Peak Signal-to-Noise Ratio}
The peak signal-to-noise ratio (PSNR)  model is a statistical measurement-based model that calculates the mean square error (MSE) for each pixel of the frame for a clip \cite{wang2003objective}. Then, the resultant MSE is used as a noise to calculate the signal-to-noise ratio. Mathematically, MSE and PSNR can be derived as follow: 
\begin{equation}
MSE = \frac{1}{MN}\sum_{i=0}^{M-1} \sum_{j=0}^{N-1} \left[I(i,j) - K(i,j)\right]^2
\end{equation}    
\par
\begin{equation}
PSNR = 20. log_{10} \left(\frac{MAX_{I}}{\sqrt{MSE}} \right) 
\end{equation}
\par
where $I$ represent the source frame with dimensions of $M\times N$ and error approximation of the source frame as $K$. 

\subsection{Structural Similarity Index Metric}
The structural similarity index matrix (SSIM) computes the clip quality based on the structural similarity (luminance,contrast, and structural comparison) between the source and distorted clip \cite{wang2004image}. SSIM can be calculated as follows:
\begin{equation}
SSIM=\frac{\left(2 \mu_{x} \mu_{y}+C_{1}\right)\left(2 \sigma_{x y}+C_{2}\right)}{\left(\mu_{x}^{2}+\mu_{y}^{2}+C_{1}\right)\left(\sigma_{x}^{2}+\sigma_{y}^{2}+C_{2}\right)}
\end{equation}
\par 
where $x$ and $y$ are the distorted and source frames from the clips, respectively.
\subsection{Multi-scale SSIM Index Metric}
The multi-scale SSIM index metric is an improved form of the SSIM metric designed to compute the quality of a  frame of clip on multiple scales \cite{wang2003multiscale}. The scale has a highest scale as $M$, and a lowest scale used for measuring luminance, while the contrast and structural comparison are measured on the $j$ scale. 
\subsection{Visual Signal-to-Noise Ratio}
Based on the near-threshold and suprathreshold features of human vision, the visual signal-to-noise ratio (VSNR) quantifies the visual fidelity within the frames of a clip. A visual fidelity greater than the thresholds is mapped to describe the clip quality of the clip \cite{chandler2007vsnr}.  
\subsection{Information Fidelity Criterion}
The information fidelity criterion (IFC) based method is a natural scene statistics (NSS) method in which the transformation of the source clip to the wavelet domain is performed by extracting the information based on NSS \cite{sheikh2005information}. The same procedure is followed for the distorted clip. Then, a single model for evaluating the visual quality of the clip is obtained by integrating the two extracted quantities. IFC can be computed as follows:
\begin{equation}
IFC=\sum_{k \in \text {subbands}} I\left(C^{N_{k}, k} ; D^{N_{k}, k} | s^{N_{k}, k}\right)
\end{equation}
\par 
where, ${C^{N_{k}, k}}$ represents the ${N_{k}}$ coefficients of the ${k_{th}}$ sub-band from the random field (RF) \cite{sheikh2005information}.
\subsection{Visual Information Fidelity}
The visual information fidelity (VIF) quantifies and extracts certain information by transforming each frame of both distorted and source clips into the wavelet domain. Both of these are based on the human visual system, and integrated to compute the distorted frame's visual quality \cite{han2013new}. 
\begin{equation}
VIF=\frac{\sum_{j \in \text {subbands}} I\left(\vec{C}^{N, j} ; \vec{F}^{N, j} | s^{N, j}\right)}{\sum_{j \in \text {subbands}} I\left(\vec{C}^{N, j} ; \vec{E}^{N, j} | s^{N, j}\right)}
\end{equation}
\par 
where the sub-bands of interest are represented as $\sum$; $E$ and $F$ represent the source and distorted frames, respectively; and $\vec{C}^{N, j}$ describes the $N$ elements of the random field ${C_{j}}$, representing the sub-band coefficients \cite{han2013new}. 
 
\begin{figure*}[t]
	\vspace{-10pt}
	\subfloat[]{
		\centering\includegraphics[width=6cm,height=5.6cm]{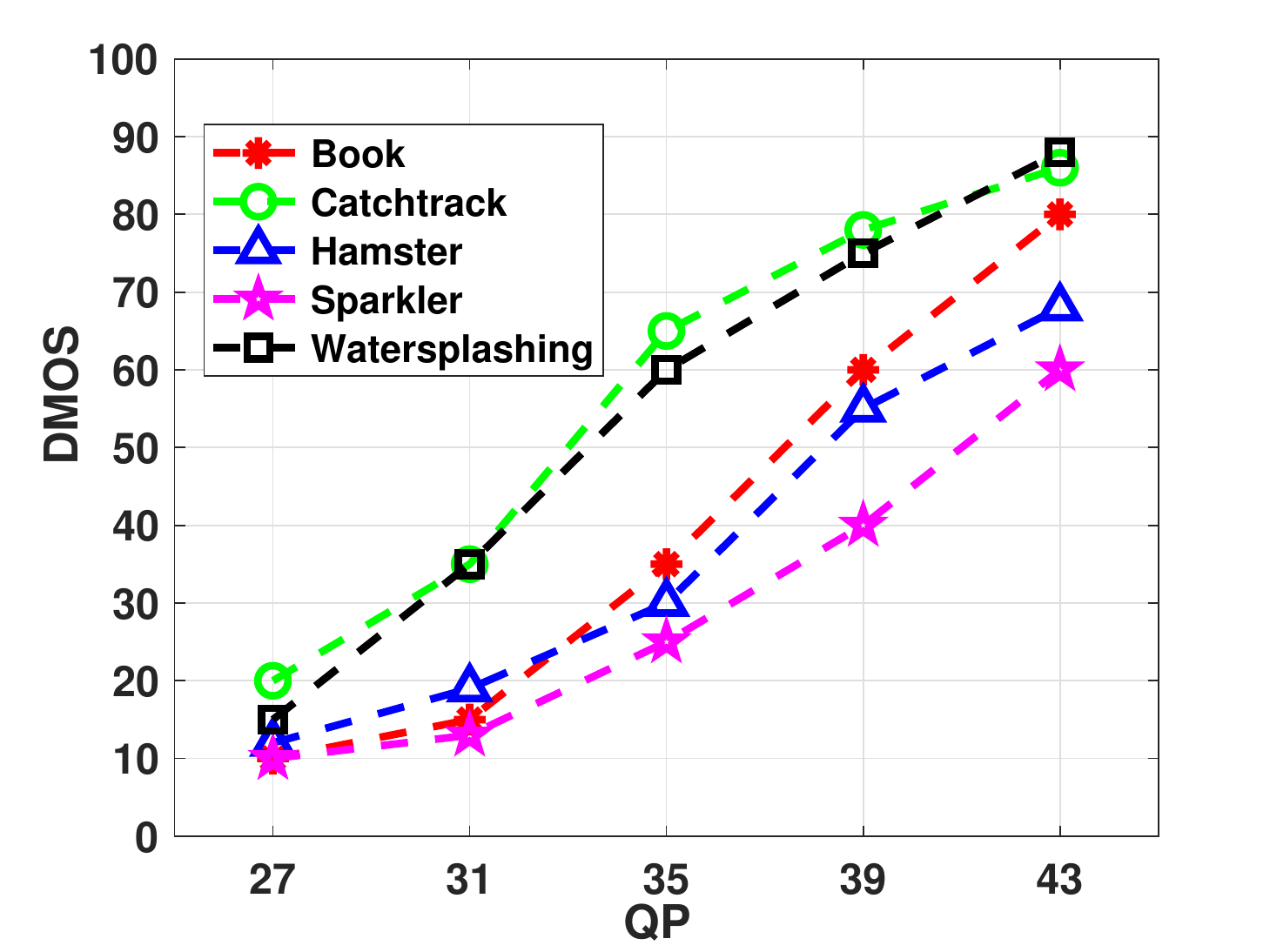}
	}
	\subfloat[]{
		\centering\includegraphics[width=6cm,height=5.5cm]{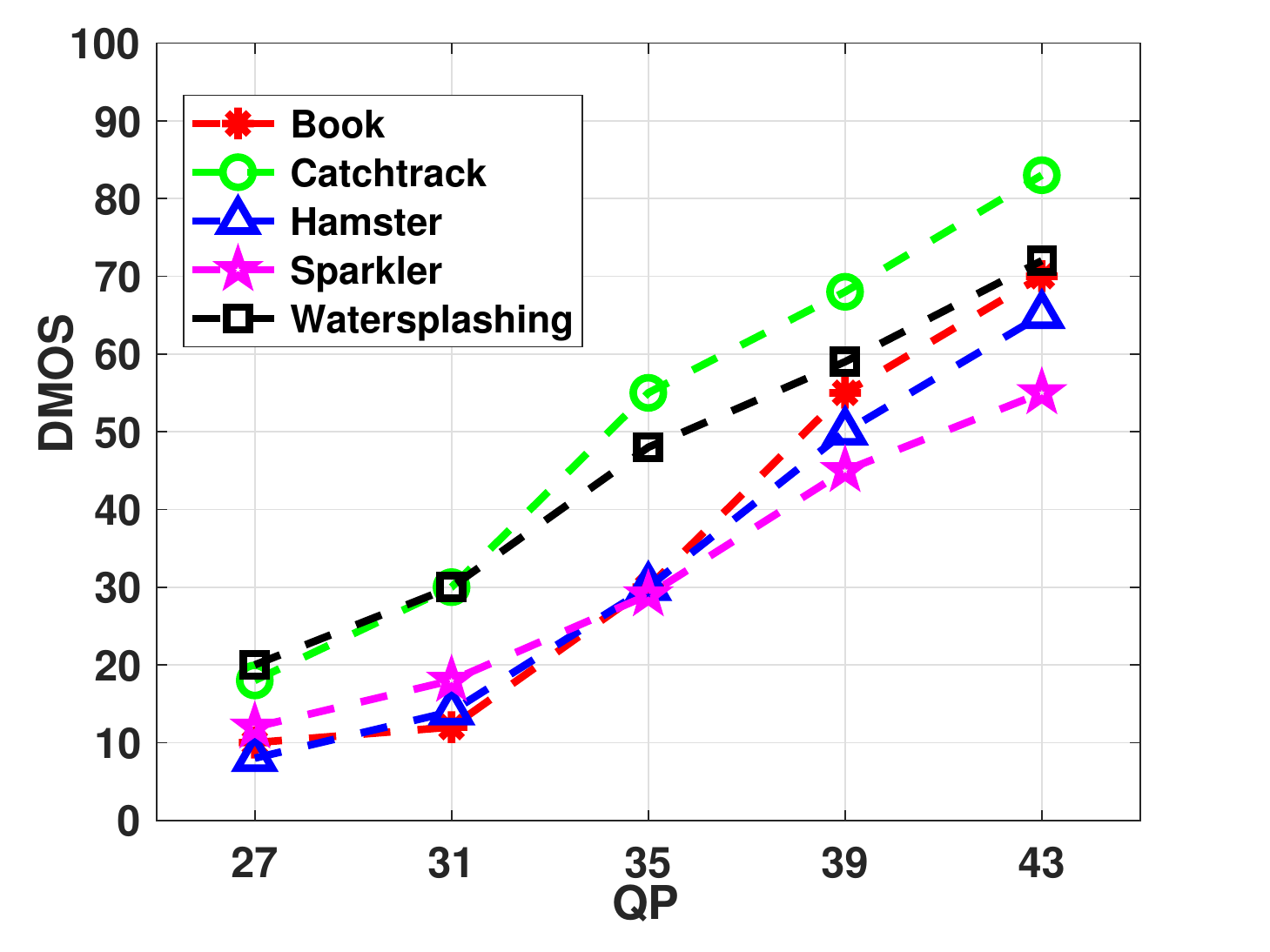}
	}
	\subfloat[]{
		\centering\includegraphics[width=6cm,height=5.6cm]{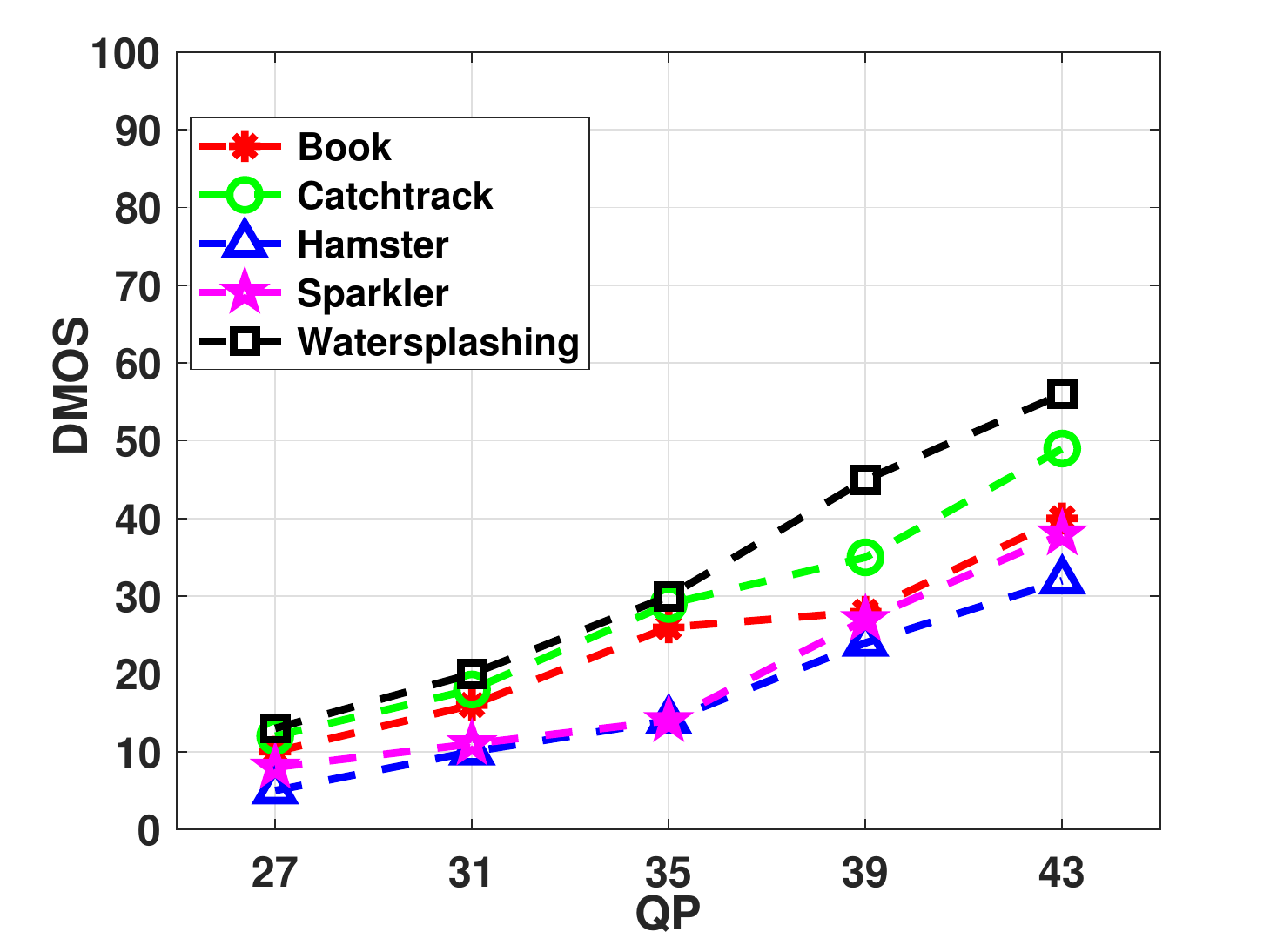}
	}
	\caption{Subjective evaluation corresponding different frame rates encoded with H.265/HEVC encoder as DMOS values against QPs levels, where (a) 15fps, (b) 30fps, and (c) 60fps.}
	\vspace{-10pt}
\end{figure*}

\begin{figure*}[t]
	\vspace{-10pt}
	\subfloat[]{
		\centering\includegraphics[width=6cm,height=5.6cm]{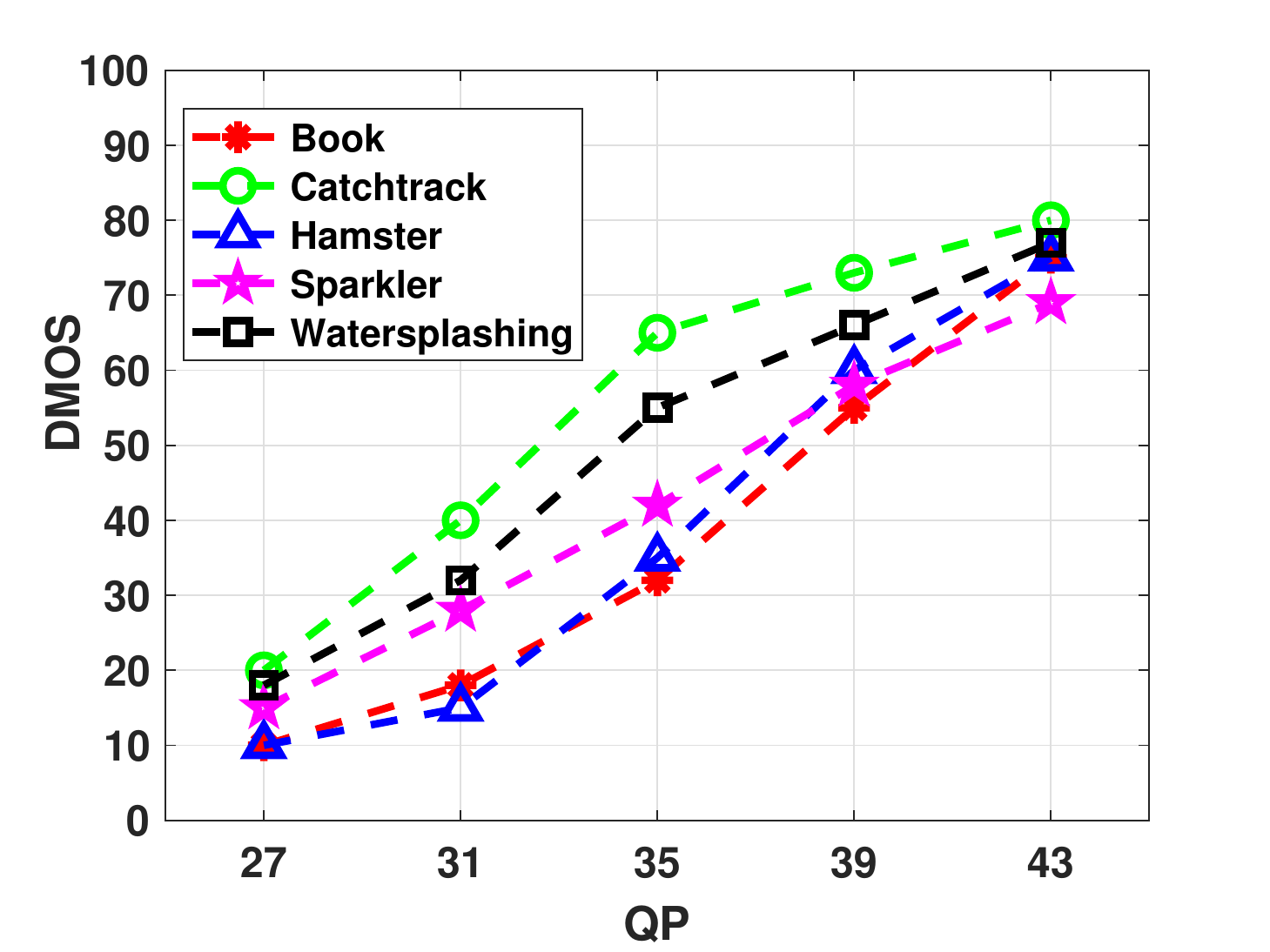}
	}
	\subfloat[]{
		\centering\includegraphics[width=6cm,height=5.5cm]{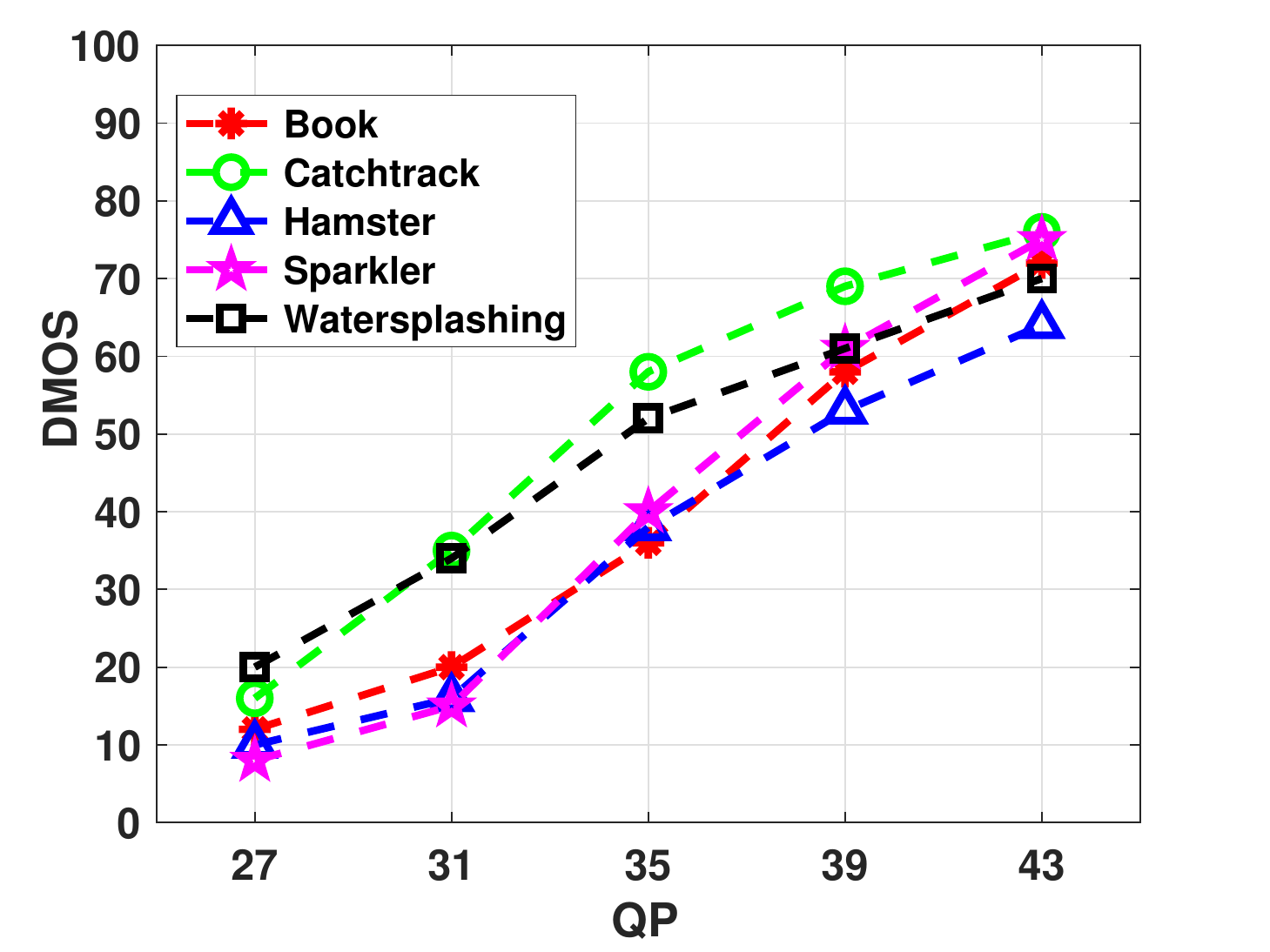}
	}
	\subfloat[]{
		\centering\includegraphics[width=6cm,height=5.6cm]{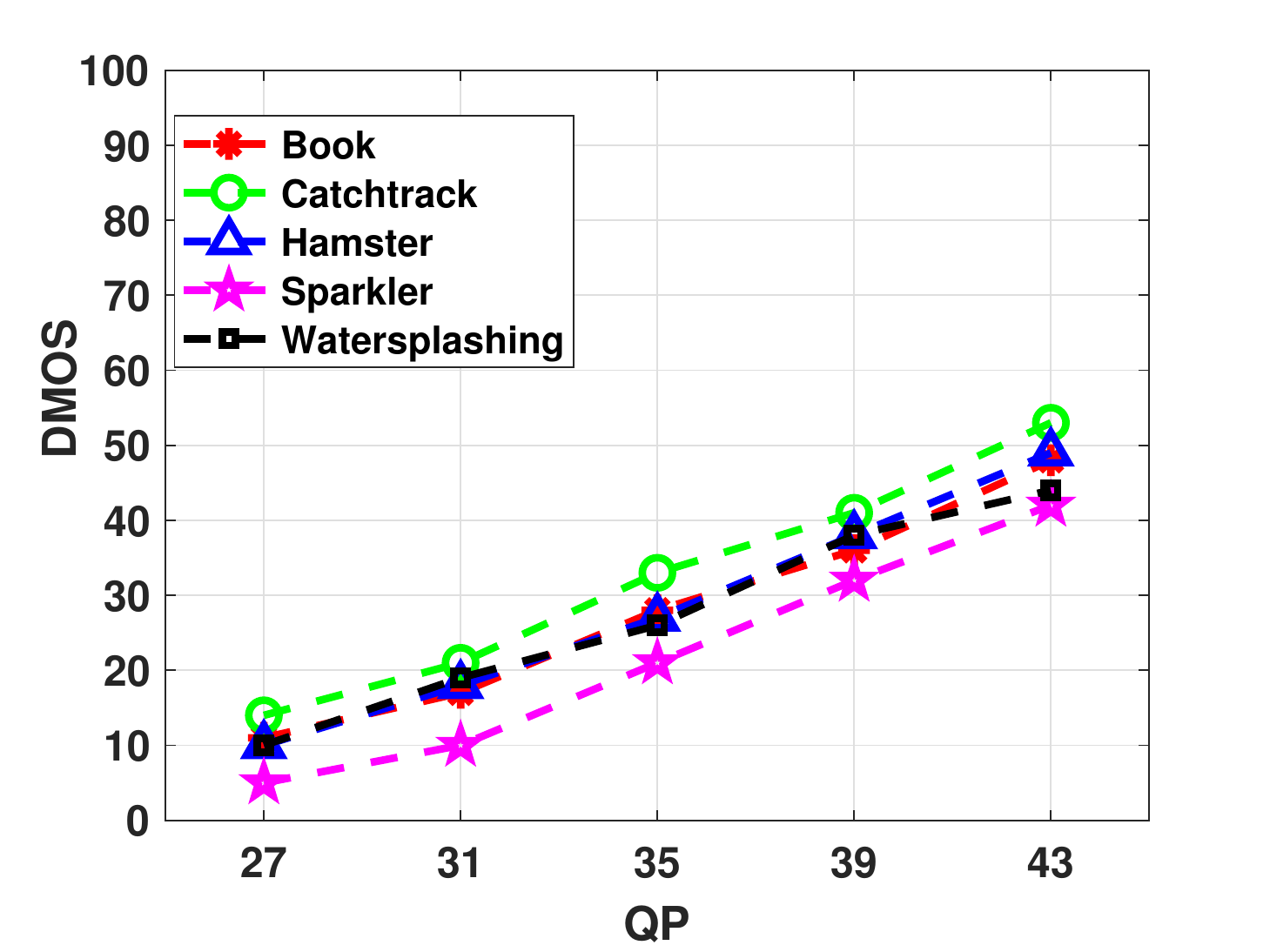}
	}
	\caption{Subjective evaluation corresponding different frame rates encoded with VP9 encoder as DMOS values against QPs levels, where (a) 15fps, (b) 30fps, and (c) 60fps.}
	\vspace{-10pt}
\end{figure*}
\subsection{Pixel-based VIF}
The pixel-based VIF is a lesser complex version of the VIF VQA metric, which extracts information on the pixel levels of the frame from both the source clip and distorted clip \cite{han2013new}.  
\subsection{Universal quality index}
This VQA metric measures the structural distortions in a clip, and then maps these measurements into a model for predicting the visual quality of the clip. In the UQI metric, the quality $Q$ can be calculated as follows: 
\begin{equation}
Q=\frac{4 \sigma_{x y} \mu_{x} \mu_{y}}{\left(\sigma_{x}^{2}+\sigma_{y}^{2}\right)\left(\mu_{x}^{2}+\mu_{y}^{2}\right)}
\end{equation}
\par 
where $x$ and $y$ are the source and distorted frames from the clips, respectively.  
\subsection{Noise Quality Measure}
This VQA metric considers the deviation in the local luminance mean, contrast sensitivity, and contrast measures of the clip \cite{damera2000image}. The noise quality measure (NQM) is a weighted SNR measure between the source and encoded clips, and can be calculated as follows:     
\begin{equation}
NQM=10 \log _{10}\left(\frac{\sum_{x} \sum_{y} O_{s}^{2}(x, y)}{\sum_{x} \sum_{y}\left(O_{s}(x, y)-I_{s}(x, y)\right)^{2}}\right)
\end{equation}
\par 
where, ${O_{s}(x,y)}$ and ${I_{s}(x,y)}$ describes the simulated versions of the restored and source frame, respectively \cite{damera2000image}. 
\subsection{Weighted Signal-to-Noise Ratio}
The VQA metric utilizes the contrast sensitivity function (CSF) that describes the weighted signal-to-noise ratio (WSNR) as a ratio of the averaged weighted signal power to the averaged weighted noise power given on a scale of dB. 
\subsection{Video Multimethod Assessment Fusion}
The video multimethod assessment fusion (VMAF)  is an FR metric that evaluates the quality of a clip based on artifacts such as compression and scaling. The VMAF employs current image quality metrics, such as VIF, Detail Loss Metric, Mean Co-Located Pixel Difference, and anti-noise signal-to-noise ratio for predicting the clip quality  \cite{netflix2019video}. 
\par 
The FR VQA metrics explained in this section are available freely online and more detail for their mathematical understandings can be found in \cite{wang2003objective} and each respective reference. In this study, we  simulated these FR metrics using the recommended simulation parameters obtained from each corresponding reference. 
\section{Results and Discussion}
\subsection{Subjective Quality Assessment Measurements}
This section describes in detail the subjective measurement for 15fps, 30fps, and 60fps clips using H.265/HEVC and VP9 encoders. For each frame rate, the subjective measurements are given as DMOS values against different QP levels for both encoders. 
\subsubsection{Subjective Test Results for H.265/HEVC}
For the subjective quality assessment, five clips (Books, Catchtrack, Hamster, Sparkler, and Watersplashing) are chosen based on the SI and TI plotting shown in Fig. 2. These five videos are encoded using an HEVC encoder, as explained in section III. A detailed subjective quality assessment for all three frame rates (15fps, 30fps, and 60fps) using the subjective perceptual video quality tool provided by MSU is conducted \cite{Anu:2013}.  
\par 
The subjective test comprised of showing the selected participants SRCs and EVSs, but belonging to the same category for example (Books). The subjective assessment is conducted for each frame rate, i.e., 15fps, 30fps, and 60fps, resulting in a total of 75 EVSs. Then the participants are asked to record their OS using DSCQS type-II, where the DMOS values by each participant, clip, and at each QP level are obtained using (1) and (2), explained in section IV. Fig. 3 shows the subjective measurements as DMOS values for all three frame rates using the HEVC encoder at each QP level. It can be seen that for all five EVSs, the DMOS values change with the increasing QP level. Higher QP levels tend to deteriorate the EVSs quality, resulting in higher DMOS values. Moreover, HFRs tend to show better quality than low frame rates. The main reason for selecting the three frame rates is to investigate the impact of different levels of compression by employing different encoders, for determining a frame rate that can reflect better user-perceived visual quality.  
\par 
In Fig. 3, for all frame rates, a uniform increase can be observed in the DMOS values as the QP increases, where a QP range of 27-31 shows a uniform behavior of DMOS values, i.e., corresponding to $\it{Excellent}$ quality for all EVs according to Table. II. Fig. 3 also shows different DMOS values for each EVSs due to different SI and TI values i.e., low to high. For example, the behaviors of Catchtrack and Watersplashing EVs show higher DMOS values for QP ranging within 35-39 at 15fps and 30fps, depicted in Figs. 3(a) and 3(b), reflecting $\it{Poor}$ quality according to Table. II. These figures also shows that the DMOS values for Catchtrack and Watersplashing deteriorate for QP ranging within 39-43 reflected as $\it{Bad}$ quality. Furthermore, at QP ranging within 35-39, the DMOS values for Hamster and Sparkling at 15fps and 30fps, as shown in Fig. 3(a) and Fig. 3(b), respectively, reflect a $\it{Good}$ quality while those for QP ranging within 39-43 reflect $\it{Fair}$ quality, according to Table. II.    
\par 
Fig. 3(c) shows that, for 60fps, even for higher QP levels ranging within 39-43, the DMOS values reflect a $\it{Fair}$ quality for all EVSs, according to Table. II. Note that EVSs such as Books, Hamster, and Sparkler show lower DMOS values for QP ranging within 27-35 as $\it{Excellent}$ quality, while those for QP ranging within 35-39 as $\it{Good}$ quality, according to Table. II and Fig. 3(c). However, when the QP level ranges within 39-43, the quality reflected in terms of DMOS values for Watersplashing is $\it{Fair}$ for 60fps, as shown in Fig. 3(c).    
\par 
From the subjective measurements generated in terms of DMOS values, as shown in Fig. 3, there is a clear impact of different frame rates at different compression (QP) levels on the user-perceived quality. HFR such as those shown in Fig. 3(c) exhibit a $\it{Good}$ visual quality even when the clips are compressed at higher QP levels. Thus, when a network has low bit-rate requirements, H.265/HEVC preserves the perceptual quality of the EVs even at the QP level ranging within 35-43. However, mixed DMOS values are reported at frame rates of 15fps and 30fps for QP level ranging within 31-39 for different EVSs, as shown in Figs. 3(b) and 3(c).  

\begin{table*}[]
	\caption{FR-VQM objective quality models corresponding each clip at 15fps for H.265/HEVC.}
	\resizebox{18.00cm}{!}{
		\renewcommand{\arraystretch}{1.1}
		\begin{tabular}{|c|c|c|c|c|c|c|c|c|c|c|c|}
			\hline
			\multirow{2}{*}{\textbf{QP levels}} & \multicolumn{11}{c|}{\textbf{FR-VQA performance metrics for EVs Books}}                                                                                                                    \\ \cline{2-12} 
			& \textbf{PSNR} & \textbf{SSIM} & \textbf{MS-SSIM} & \textbf{VSNR} & \textbf{WSNR} & \textbf{NQM} & \textbf{UQI} & \textbf{VIF} & \textbf{VIFP} & \textbf{IFC} & \textbf{VMAF} \\ \hline
			\textbf{27}                         & \textbf{38.8320}       & \textbf{0.9248 }       & 0.9106           & 34.6771       & \textbf{32.5068}       & 32.8726      & \textbf{0.7668}      & \textbf{0.4587}      & 0.5831        & 2.2473       & \textbf{84.3624}      \\ \hline
			\textbf{31}                         & \textbf{34.4770}       & \textbf{0.9061}      & 0.8924           & 32.6094       & \textbf{30.5883}       & 30.1283      & \textbf{0.7381}       & \textbf{0.3984}       & 0.4810        & 2.1654       & \textbf{82.7945}     \\ \hline
			\textbf{35}                         & 31.7791       & \textbf{0.9161}        & 0.8815           & 31.2886       & 29.6138       & 28.2058      & \textbf{0.6216}       & 0.3107       & 0.4121        & 1.7286       & \textbf{78.8201}       \\ \hline
			\textbf{39}                         & 29.6824       & 0.8472        & 0.9482           & 31.0826       & 29.2311       & 27.4281      & 0.5264       & 0.2882       & 0.3981        & 1.4681       & 75.8801       \\ \hline
			\textbf{43}                         & 26.7703       & 0.7741        & 0.9224           & 28.0772       & 26.8163       & 23.1633      & 0.4775       & 0.2301       & 0.3402        & 1.1006       & 73.0758       \\ \hline
			
			\multicolumn{12}{c}{(a)} \\ \hline

			\multirow{2}{*}{\textbf{QP levels}} & \multicolumn{11}{c|}{\textbf{FR-VQA performance metrics for EVs CatchTrack}}                                                                                                                    \\ \cline{2-12} 
			& \textbf{PSNR} & \textbf{SSIM} & \textbf{MS-SSIM} & \textbf{VSNR} & \textbf{WSNR} & \textbf{NQM} & \textbf{UQI} & \textbf{VIF} & \textbf{VIFP} & \textbf{IFC} & \textbf{VMAF} \\ \hline
			\textbf{27}                         & \textbf{37.2140 }      & \textbf{0.8681}        & 0.9083           & 32.5904       & \textbf{31.7114 }      & 31.0041      & \textbf{0.7513}       & \textbf{0.4008}     & 0.5125        & 2.2018       & \textbf{79.0483}      \\ \hline
			\textbf{31}                         & \textbf{34.8260}       & \textbf{0.8215}        & 0.8810           & 31.4811       & \textbf{29.8468}       & 29.6022      & \textbf{0.7220}       & \textbf{0.3126}      & 0.4226        & 1.8066       & \textbf{77.2097}       \\ \hline
			\textbf{35}                         & 28.5797       & 0.9435       & 0.8907           & 30.6371       & 28.2073       & 27.0447      & 0.5708       & 0.2986       & 0.3846        & 1.6428       & \textbf{76.4552}       \\ \hline
			\textbf{39}                         & 27.8335       & 0.7894        & 0.9011           & 28.8846       & 26.2462       & 25.1035      & 0.4902       & 0.2364       & 0.3406        & 1.3046       & 74.2063       \\ \hline
			\textbf{43}                         & 24.8725       & 0.7137        & 0.9047           & 25.6643       & 24.2487       & 21.7748      & 0.4261       & 0.2218       & 0.3384        & 0.9838       & 72.4018       \\ \hline   
			
			\multicolumn{12}{c}{(b)} \\ \hline

			\multirow{2}{*}{\textbf{QP levels}} & \multicolumn{11}{c|}{\textbf{FR-VQA performance metrics for EVs Hamster}}                                                                                                \\ \cline{2-12} 
			& \textbf{PSNR} & \textbf{SSIM} & \textbf{MS-SSIM} & \textbf{VSNR} & \textbf{WSNR} & \textbf{NQM} & \textbf{UQI} & \textbf{VIF} & \textbf{VIFP} & \textbf{IFC} & \textbf{VMAF} \\ \hline
			\textbf{27}                         & \textbf{38.2380   }    & \textbf{0.9203}        & 0.9401           & 34.4263       & \textbf{32.3725 }      & 31.4683      & \textbf{0.7736 }      & \textbf{0.4289}       & 0.5337        & 2.1551       & \textbf{84.1636}       \\ \hline
			\textbf{31}                         & \textbf{36.0250  }     & \textbf{0.8756}        & 0.9286           & 32.2073       & \textbf{30.4031}       & 30.0047      & \textbf{0.7301}       & \textbf{0.3842}     & 0.4791        & 2.1351       & \textbf{81.2389}      \\ \hline
			\textbf{35}                         & \textbf{31.2063}      & 0.8701        & 0.8240           & 31.0025       & \textbf{29.4008}      & 28.0066      & \textbf{0.6030}       & 0.3001       & 0.4062        & 1.7113       & \textbf{79.2841 }      \\ \hline
			\textbf{39}                         & 28.3801       & 0.8035        & 0.8557           & 28.8038       & 26.2157       & 25.8868      & \textbf{0.5008}       & 0.2563       & 0.3597        & 1.3882       & \textbf{75.0375}    \\ \hline
			\textbf{43}                         & 26.0382       & 0.7581        & 0.8815           & 25.9011       & 25.6193       & 22.6891      & 0.4530       & 0.2103       & 0.3255        & 0.8836       & 73.8921       \\ \hline
			
			\multicolumn{12}{c}{(c)} \\ \hline
			\multirow{2}{*}{\textbf{QP levels}} & \multicolumn{11}{c|}{\textbf{FR-VQA performance metrics for EVs Sparkler}}                                                                                                \\ \cline{2-12} 
			& \textbf{PSNR} & \textbf{SSIM} & \textbf{MS-SSIM} & \textbf{VSNR} & \textbf{WSNR} & \textbf{NQM} & \textbf{UQI} & \textbf{VIF} & \textbf{VIFP} & \textbf{IFC} & \textbf{VMAF} \\ \hline
			\textbf{27}                         & \textbf{38.6280 }      & \textbf{0.9463}        & 0.9285           & 34.5804       & \textbf{32.4886}       & 32.7067      & \textbf{0.7742}       & \textbf{0.4503}       & 0.5801        & 2.2406       & \textbf{83.9842}        \\ \hline
			\textbf{31}                         & \textbf{36.5070}       & \textbf{0.9124}        & 0.9315           & 32.6481       & \textbf{30.6682}       & 30.1771      & \textbf{0.7384}       & \textbf{0.3987}       & 0.4935        & 2.1668       & \textbf{83.0482}       \\ \hline
			\textbf{35}                         & \textbf{31.8759}      & \textbf{0.9064}        & 0.8804           & 31.4685       & 29.8842       & 28.2461      & \textbf{0.6184}       & 0.3187       & 0.4188        & 1.7301       & \textbf{77.9120}       \\ \hline
			\textbf{39}                         & 29.8736       & 0.8581        & 0.8801           & 31.1386       & 29.2846       & 27.5907      & 0.5410       & 0.2901       & 0.4013        & 1.6684       & \textbf{76.0103 }      \\ \hline
			\textbf{43}                         & 26.5509       & 0.7885        & 0.9004           & 28.0425       & 26.7246       & 26.7061      & 0.4568       & 0.2201       & 0.3317        & 0.9802       & 74.1601       \\ \hline
			
			\multicolumn{12}{c}{(c)} \\ \hline
			\multirow{2}{*}{\textbf{QP levels}} & \multicolumn{11}{c|}{\textbf{FR-VQA performance metrics for EVs Watersplashing}}                                                                                                \\ \cline{2-12}
			& \textbf{PSNR} & \textbf{SSIM} & \textbf{MS-SSIM} & \textbf{VSNR} & \textbf{WSNR} & \textbf{NQM} & \textbf{UQI} & \textbf{VIF} & \textbf{VIFP} & \textbf{IFC} & \textbf{VMAF} \\ \hline
			\textbf{27}                         & \textbf{37.9580}       & \textbf{0.8876}        & 0.8177           & 32.8862       & \textbf{31.8903}       & 31.2511      & \textbf{0.7540}      & \textbf{0.4106}      & 0.5104        & 2.2107       & \textbf{81.4425}       \\ \hline
			\textbf{31}                         & \textbf{35.2070}       & \textbf{0.8829}        & 0.8481           & 31.8375       & \textbf{30.0047}      & 29.8904      & \textbf{0.7266}       & \textbf{0.3473}       & 0.4468        & 1.8972       & \textbf{79.6701}       \\ \hline
			\textbf{35}                         & 29.3205       & \textbf{0.8371}        & 0.8481           & 30.9404       & 28.7905       & 27.6603      & \textbf{0.5817}      & 0.3103       & 0.4103        & 1.7294       & \textbf{78.0137}       \\ \hline
			\textbf{39}                         & 27.5791       & 0.8037        & 0.8277           & 28.7168       & 26.2005       & 24.6815      & 0.4803       & 0.2305       & 0.3387        & 1.2576       & \textbf{75.8371}       \\ \hline
			\textbf{43}                         & 24.2571       & 0.7102        & 0.8604           & 25.4618       & 24.2176       & 21.3118      & 0.4208       & 0.2204       & 0.3319        & 0.988        & 73.1003       \\ \hline
			
			\multicolumn{12}{c}{(e)} \\ 
		\end{tabular}
	}
\end{table*}   
\subsubsection{Subjective Test Results for VP9}
Similarly, for subjective quality assessment, five clips (Books, Catchtrack, Hamster, Sparkler, and Watersplashing) are chosen based on the SI and TI plotting shown in Fig. 2, and encoded using the VP9 encoder, as explained in section III. A detailed subjective quality assessment for all three frame rates is conducted using the subjective perceptual video quality tool provided by MSU \cite{Anu:2013}.  

\begin{table*}[t]
	\caption{FR-VQM objective quality models corresponding each clip at 30fps for H.265/HEVC.}
	\resizebox{18.00cm}{!}{
		\renewcommand{\arraystretch}{1.1}
		\begin{tabular}{|c|c|c|c|c|c|c|c|c|c|c|c|}
			\hline
			\multirow{2}{*}{\textbf{QP levels}} & \multicolumn{11}{c|}{\textbf{FR-VQA performance metrics for EVs Books}}                                                                                                            \\ \cline{2-12} 
			& \textbf{PSNR} & \textbf{SSIM} & \textbf{MS-SSIM} & \textbf{VSNR} & \textbf{WSNR} & \textbf{NQM} & \textbf{UQI} & \textbf{VIF} & \textbf{VIFP} & \textbf{IFC} & \textbf{VMAF} \\ \hline
			\textbf{27}                         & \textbf{44.7351 }      & \textbf{0.9441}        & 0.9485           & 41.8217       & \textbf{40.0285}       & 37.8825      & \textbf{0.7906}       & \textbf{0.6824}       & 0.7425        & 2.8847       & \textbf{85.5071}       \\ \hline
			\textbf{31}                         & \textbf{42.0081}       & \textbf{0.9488}      & 0.9551           & 39.2651       & \textbf{36.5881}      & 35.6603      & \textbf{0.7601}      & \textbf{0.4818}      & 0.5824        & 2.4866       & \textbf{83.8816}      \\ \hline
			\textbf{35}                         & \textbf{38.8905}       & \textbf{0.9204}        & 0.9351           & 36.8091       & \textbf{33.6153}       & 31.2607      & \textbf{0.7588}    & 0.4486       & 0.5231        & 2.1246       & \textbf{78.8020}       \\ \hline
			\textbf{39}                         & \textbf{35.6207}      & \textbf{0.9181}        & 0.9535           & 34.8074       & \textbf{31.2885}       & 27.8083      & \textbf{0.6802}       & 0.3891       & 0.4682        & 1.8841       & \textbf{77.5408}       \\ \hline
			\textbf{43}                         & 34.4066       & 0.8974        & 0.9330           & 31.6581       & 28.5886       & 24.9041      & 0.6180       & 0.3266       & 0.4001        & 1.3882       & 75.8209       \\ \hline		
			\multicolumn{12}{c}{(a)} \\ \hline
			
			\multirow{2}{*}{\textbf{QP levels}} & \multicolumn{11}{c|}{\textbf{FR-VQA performance metrics for EVs Catchtrack}}                                                                                                       \\ \cline{2-12} 
			& \textbf{PSNR} & \textbf{SSIM} & \textbf{MS-SSIM} & \textbf{VSNR} & \textbf{WSNR} & \textbf{NQM} & \textbf{UQI} & \textbf{VIF} & \textbf{VIFP} & \textbf{IFC} & \textbf{VMAF} \\ \hline
			\textbf{27}                         & \textbf{43.8270}      & \textbf{0.9404}        & 0.9506           & 40.6846       & \textbf{39.2481}       & 36.8746      & \textbf{0.7830}      & \textbf{0.5881}      & 0.6583        & 2.5694       & \textbf{82.1823}      \\ \hline
			\textbf{31}                         & \textbf{43.4105}       & \textbf{0.9451}        & 0.9602           & 40.1506       & \textbf{38.7366}       & 36.2842      & \textbf{0.7628}      & \textbf{0.5206}      & 0.6083        & 2.5001       & \textbf{80.5022}
			\\ \hline
			\textbf{35}                         & \textbf{38.0027}       & \textbf{0.9010}        & 0.9113           & 36.1284       & \textbf{32.8251}       & 31.3582      & \textbf{0.7417}      & 0.4001       & 0.4947        & 2.0096       & \textbf{79.3301}      \\ \hline
			\textbf{39}                         & \textbf{35.4661}       & \textbf{0.9053}        & 0.9020           & 34.6627       & \textbf{31.0758}       & 27.6621      & \textbf{0.6714}       & 0.3718       & 0.4467        & 1.7648       & \textbf{77.8207}       \\ \hline
			\textbf{43}                         & 31.9801       & 0.8693        & 0.9281           & 30.8807       & 27.2691       & 24.0026      & 0.6036       & 0.3008       & 0.3861        & 1.3006       & 75.2077       \\ \hline	
			\multicolumn{12}{c}{(b)} \\ \hline
			\multirow{2}{*}{\textbf{QP levels}} & \multicolumn{11}{c|}{\textbf{FR-VQA performance metrics for EVs Hamster}}                                                                                                          \\ \cline{2-12} 
			& \textbf{PSNR} & \textbf{SSIM} & \textbf{MS-SSIM} & \textbf{VSNR} & \textbf{WSNR} & \textbf{NQM} & \textbf{UQI} & \textbf{VIF} & \textbf{VIFP} & \textbf{IFC} & \textbf{VMAF} \\ \hline
			\textbf{27}                         & \textbf{44.8801}      & \textbf{0.9382}        & 0.9221           & 41.8861       & \textbf{40.1206}       & 37.8903      & \textbf{0.7955}      & \textbf{0.6907}       & 0.7551        & 2.8986       & \textbf{84.5507}       \\ \hline
			\textbf{31}                         & \textbf{44.7022}       & \textbf{0.9201}        & 0.9441           & 41.3572       & \textbf{38.8901}       & 36.2117      & \textbf{0.7731}       & \textbf{0.6107}       & 0.7116        & 2.6204       & \textbf{82.0184}       \\ \hline
			\textbf{35}                         & \textbf{38.4103}       & \textbf{0.8642}        & 0.8810           & 36.2623       & \textbf{33.0570}       & 31.1005      & \textbf{0.7550}       & \textbf{0.4411}       & 0.5186        & 2.1103       & \textbf{80.1040}      \\ \hline
			\textbf{39}                         & \textbf{36.5810}       & \textbf{0.8860}        & 0.9433           & 34.8006       & 31.2061       & 27.7115      & 0.6988       & \textbf{0.4001}       & 0.5014        & 1.9033       & \textbf{80.5580}       \\ \hline
			\textbf{43}                         & 32.4116       & 0.8724        & 0.8950           & 31.8903       & 28.8632       & 24.9158      & 0.6204       & 0.3381       & 0.4082        & 1.3991       & 74.6200       \\ \hline
			\multicolumn{12}{c}{(c)} \\ \hline
			\multirow{2}{*}{\textbf{QP levels}} & \multicolumn{11}{c|}{\textbf{FR-VQA performance metrics for EVs Sparkler}}                                                                                                         \\ \cline{2-12} 
			& \textbf{PSNR} & \textbf{SSIM} & \textbf{MS-SSIM} & \textbf{VSNR} & \textbf{WSNR} & \textbf{NQM} & \textbf{UQI} & \textbf{VIF} & \textbf{VIFP} & \textbf{IFC} & \textbf{VMAF} \\ \hline
			\textbf{27}                         & \textbf{44.6682}       & \textbf{0.9388}       & 0.9117           & 41.7053       & \textbf{40.1133}       & 37.6041      & \textbf{0.7920}       & \textbf{0.6713}       & 0.7488        & 2.7691       & \textbf{82.6610}       \\ \hline
			\textbf{31}                         & \textbf{42.8846}       & \textbf{0.9422}        & 0.9011           & 39.8807       & \textbf{36.8688}       & 34.8005      & \textbf{0.7610}       & \textbf{0.5131}       & 0.5907        & 2.4911       & \textbf{84.5048}       \\ \hline
			\textbf{35}                         & \textbf{38.2480}       & \textbf{0.8711}        & 0.8947           & 35.8847       & \textbf{32.5611}       & 31.1001      & \textbf{0.7480}       & \textbf{0.4426}       & 0.5218        & 2.1185       & \textbf{77.2380}       \\ \hline
			\textbf{39}                         & \textbf{35.8027}       & \textbf{0.9104}        & 0.9110           & 34.8964       & 31.4662       & 27.8664      & \textbf{0.6840}       & 0.3903       & 0.4883        & 1.8241       & \textbf{76.8064}       \\ \hline
			\textbf{43}                         & 32.1184      & 0.8801        & 0.9001           & 31.6695       & 28.6064       & 28.6027      & \textbf{0.6107}       & 0.3205       & 0.3928        & 1.3258       & \textbf{75.0831}       \\ \hline	
			\multicolumn{12}{c}{(d)} \\ \hline	
			\multirow{2}{*}{\textbf{QP levels}} & \multicolumn{11}{c|}{\textbf{FR-VQA performance metrics for EVs Watersplashing}}                                                                                                   \\ \cline{2-12} 
			& \textbf{PSNR} & \textbf{SSIM} & \textbf{MS-SSIM} & \textbf{VSNR} & \textbf{WSNR} & \textbf{NQM} & \textbf{UQI} & \textbf{VIF} & \textbf{VIFP} & \textbf{IFC} & \textbf{VMAF} \\ \hline
			\textbf{27}                         & \textbf{43.5507}       & \textbf{0.9016}        & 0.8824           & 40.5114       & \textbf{39.0158}       & 39.0070      & \textbf{0.7758}       & \textbf{0.5346}       & 0.6125        & 2.5062       & \textbf{82.3358}       \\ \hline
			\textbf{31}                         & \textbf{43.5084}       & \textbf{0.9001}        & 0.9346           & 40.1664       & \textbf{37.0083}       & 36.3062      & \textbf{0.7633}       & \textbf{0.5288}       & 0.6115        & 2.5103       & \textbf{83.1003}       \\ \hline
			\textbf{35}                         & \textbf{37.7461}       & \textbf{0.8603}        & 0.9280           & 35.2005       & 32.0826       & 30.8914      & \textbf{0.7166}       & 0.4037       & 0.4988        & 2.0188       & \textbf{78.0241}       \\ \hline
			\textbf{39}                         & \textbf{34.6792}       & \textbf{0.9003}        & 0.9426           & 33.8941       & 30.7448       & 27.0027      & \textbf{0.6427}       & 0.3436       & 0.4105        & 1.6094       & \textbf{79.3350}       \\ \hline
			\textbf{43}                         & 32.0085       & 0.8810        & 0.9057           & 31.2628       & 28.1338       & 24.3361      & 0.6110       & 0.3117       & 0.3905        & 1.3106       & 74.3806       \\ \hline
			\multicolumn{12}{c}{(e)} \\ 
		\end{tabular}
	}
\end{table*}
\par 
Fig. 4 depicts the subjective measurements as DMOS values against each QP level, using the VP9 encoder for  15fps, 30fps, and 60fps clips. The subjective assessment is conducted for each frame rate clip, resulting in 75 EVSs in total. In Fig. 4, for all frame rates, a uniform increase can be observed in the DMOS values as the QP increases,  where a QP range of 27-31 shows a uniform behavior of DMOS values, i.e., corresponding to $\it{Excellent}$ quality for all EVSs, according to Table. II. However, different DMOS values are recorded for each EVSs. For example, the behaviors of Catchtrack and Watersplashing EVSs depicted in Figs. 4(a) and 4(b) at 15fps and 30fps, respectively, with QP ranging 35-39 result in $\it{Poor}$ and $\it{Fair}$ qualities, respectively, according to Table. II. Note that EVSs such as Books, Hamster, and Sparkler at 15fps and 30fps as shown in Figs. 4(a) and 4(b), for QP ranging within 31-35 reflect $\it{Good}$ quality, according to Table. II. For EVSs at 60fps, as shown in Fig. 4(c), within QP ranging 35-43, all EVs reflect $\it{Good}$ quality, except for Catchtrack which reflect $\it{Fair}$ quality, according to Table. II. However, for QP ranging within 27-31, all EVSs reflect $\it{Excellent}$ quality, according to Table.II and shown in Fig. 4(c).   
\par 
From the subjective measurements generated in terms of DMOS values, as shown in Fig. 4, we can interpret a ckear impact of different frame rates at different compression (QP) levels on the user-perceived quality. HFR such as those shown in Fig. 4(c) exhibit a $\it{Good}$ visual quality even when the clips are compressed at QP levels within 35-43. Thus, when a network has low bit-rate requirements, VP9 preserves the perceptual quality of the EVs even at QP levels within 35-43. However, mixed DMOS values are reported at frame rates of 15fps and 30fps for QP level ranging within 31-39 for different EVs, as shown in Figs. 4(b) and 4(c).
\subsection{Performance of Objective Models}
This section quantifies the relation required to be fitted between the VQA and the subjective measurements generated in terms of DMOS values explained in sections V and VI, respectively.  
\subsubsection{FR Objective VQA Results for H.265/HEVC}
Considering the 11 state-of-the-art FR-objective VQA metrics discussed in section V, including recently adopted VAMF trained on NETFLIX's videos \cite{netflix2019video}, we attempt to determine the relation between subjective and objective measurements for benchmarking \cite{wang2017video}. The detailed performance of the selected FR-VQA at all three frame rates using HEVC is listed in Table. III, Table. IV, and Table. V, respectively.   
\par 
Table. III shows the performance of HEVC using FR-VQA metrics at 15fps for different EVSs at different QP levels. It can be seen that, for each EVSs, as the QP level increases, the FR-VQA metrics vary correspondingly just like the subjective quality assessment measurements discussed in section VI(A). Due to the larger size of the coding tree unit, coding tree block, and an improved variable-block-size segmentation \cite{sullivan2012overview} in HEVC, the FR-VQA metrics such as PSNR, SSIM, WSNR, UQI, VIF, and VMAF \cite{usman2019suitability} show a better performance. However, just like different DMOS values generated for different EVSs as shown in Fig. 3(a), the performance of FR objective VQA metrics for different EVSs show different results. For QP ranging within 27-31, all EVs are showing higher values of PSNR, SSIM, WSNR, UQI, VIF, and VMAF reflected as $\it{Excellent}$ quality. Similarly, as seen in Tables. III(b) and III(e), the FR-VQA metrics in QP ranging 39-43 show lower values for FR-VQA metrics reflected as $\it{Bad}$ quality. On the other hand for QP ranging within 39-43, in Tables. III(a), III(c), and III(d), a better performance of FR-VQA metrics is reflected as a $\it{Fair}$ visual quality. These results are highly correlated with the subjective measurements shown in Fig. 3(a). Table. III also shows that for a network with low bit-rate requirements at 15fps EVSs, an acceptable quality is achieved using HEVC ranges within 27-31 based on both subjective and objective models.

\begin{table*}[]
	\caption{FR-VQM objective quality models corresponding each clip at 60fps for H.265/HEVC.}
	\resizebox{18.00cm}{!}{
		\renewcommand{\arraystretch}{1.1}
		\centering
		\begin{tabular}{|c|c|c|c|c|c|c|c|c|c|c|c|}
			\hline
			\multirow{2}{*}{\textbf{QP levels}} & \multicolumn{11}{c|}{\textbf{FR-VQA performance metrics for EVs Books}}                                                                                                            \\ \cline{2-12} 
			& \textbf{PSNR} & \textbf{SSIM} & \textbf{MS-SSIM} & \textbf{VSNR} & \textbf{WSNR} & \textbf{NQM} & \textbf{UQI} & \textbf{VIF} & \textbf{VIFP} & \textbf{IFC} & \textbf{VMAF} \\ \hline
			\textbf{27}                         & 48.8261       & 0.9620        & 0.9815           & 45.8467       & 43.8864       & 41.7906      & 0.8516       & 0.7813       & 0.8847        & 3.5952       & 86.6210       \\ \hline
			\textbf{31}                         & 44.0517       & 0.9483        & 0.9772           & 42.0835       & 39.1624       & 38.0012      & 0.7813       & 0.6248       & 0.7402        & 2.8816       & 84.5813       \\ \hline
			\textbf{35}                         & 41.2810       & 0.9325        & 0.9423           & 39.7924       & 36.2206       & 35.8472      & 0.7791       & 0.5816       & 0.6824        & 2.5653       & 81.8371       \\ \hline
			\textbf{39}                         & 37.6040       & 0.9131        & 0.9324           & 36.6473       & 33.2874       & 30.6247      & 0.7337       & 0.4882       & 0.5894        & 2.1003       & 78.9040       \\ \hline
			\textbf{43}                         & 34.8806       & 0.9064        & 0.9388           & 34.2641       & 31.6682       & 28.0281      & 0.6817       & 0.4286       & 0.5066        & 1.7885       & 76.3104       \\ \hline
			\multicolumn{12}{c}{(a)} \\ \hline	
			\multirow{2}{*}{\textbf{QP levels}} & \multicolumn{11}{c|}{\textbf{FR-VQA performance metrics for EVs Catchtrack}}                                                                                                       \\ \cline{2-12} 
			& \textbf{PSNR} & \textbf{SSIM} & \textbf{MS-SSIM} & \textbf{VSNR} & \textbf{WSNR} & \textbf{NQM} & \textbf{UQI} & \textbf{VIF} & \textbf{VIFP} & \textbf{IFC} & \textbf{VMAF} \\ \hline
			\textbf{27}                         & 48.0377       & 0.9581        & 0.9601           & 44.8823       & 42.0161       & 41.0035      & 0.8003       & 0.7475       & 0.8553        & 3.4026       & 86.0843       \\ \hline
			\textbf{31}                         & 45.8220       & 0.9517        & 0.9706           & 43.2258       & 40.2688       & 38.8825      & 0.7968       & 0.6681       & 0.7618        & 3.0894       & 84.2351       \\ \hline
			\textbf{35}                         & 41.1104       & 0.9310        & 0.9516           & 39.5041       & 36.1008       & 35.6483      & 0.7440       & 0.5722       & 0.6611        & 2.5281       & 81.8711       \\ \hline
			\textbf{39}                         & 36.6813       & 0.9140        & 0.9386           & 37.2681       & 34.4661       & 31.3682      & 0.7582       & 0.5081       & 0.6013        & 2.2911       & 79.3581       \\ \hline
			\textbf{43}                         & 35.2214       & 0.9031        & 0.9360           & 34.8564       & 31.8677       & 28.4628      & 0.7106       & 0.4663       & 0.5527        & 1.9762       & 76.6558       \\ \hline	
			\multicolumn{12}{c}{(b)} \\ \hline
			\multirow{2}{*}{\textbf{QP levels}} & \multicolumn{11}{c|}{\textbf{FR-VQA performance metrics for EVs Hamster}}                                                                                                          \\ \cline{2-12} 
			& \textbf{PSNR} & \textbf{SSIM} & \textbf{MS-SSIM} & \textbf{VSNR} & \textbf{WSNR} & \textbf{NQM} & \textbf{UQI} & \textbf{VIF} & \textbf{VIFP} & \textbf{IFC} & \textbf{VMAF} \\ \hline
			\textbf{27}                         & 48.5519       & 0.9480        & 0.9530           & 45.2653       & 43.1863       & 41.4342      & 0.8210       & 0.7701       & 0.8607        & 3.5204       & 84.8837       \\ \hline
			\textbf{31}                         & 45.1370       & 0.9338        & 0.9211           & 43.8748       & 40.6236       & 38.2401      & 0.7915       & 0.6463       & 0.7511        & 2.9862       & 83.2093       \\ \hline
			\textbf{35}                         & 40.5612       & 0.9082        & 0.9311           & 38.6883       & 35.3162       & 34.2682      & 0.7701       & 0.5403       & 0.6358        & 2.4866       & 80.5504       \\ \hline
			\textbf{39}                         & 38.5507       & 0.9088        & 0.9377           & 37.2016       & 34.3004       & 31.3256      & 0.7548       & 0.5006       & 0.6007        & 2.2135       & 79.0370       \\ \hline
			\textbf{43}                         & 34.8620       & 0.8907        & 0.9315           & 34.2304       & 31.5064       & 28.0232      & 0.6720       & 0.4297       & 0.4893        & 1.7904       & 75.2610       \\ \hline		
			\multicolumn{12}{c}{(c)} \\ \hline		
			\multirow{2}{*}{\textbf{QP levels}} & \multicolumn{11}{c|}{\textbf{FR-VQA performance metrics for EVs  Sparkler}}                                                                                                         \\ \cline{2-12} 
			& \textbf{PSNR} & \textbf{SSIM} & \textbf{MS-SSIM} & \textbf{VSNR} & \textbf{WSNR} & \textbf{NQM} & \textbf{UQI} & \textbf{VIF} & \textbf{VIFP} & \textbf{IFC} & \textbf{VMAF} \\ \hline
			\textbf{27}                         & 48.8805       & 0.9553        & 0.9684           & 46.0467       & 43.9002       & 41.9004      & 0.8614       & 0.7884       & 0.8878        & 3.5611       & 85.9020       \\ \hline
			\textbf{31}                         & 45.6641       & 0.9452        & 0.9448           & 43.9507       & 40.8448       & 38.6291      & 0.7940       & 0.6604       & 0.7384        & 3.0015       & 84.0014       \\ \hline
			\textbf{35}                         & 41.0370       & 0.8922        & 0.9084           & 39.4807       & 36.1001       & 35.6031      & 0.7725       & 0.5701       & 0.6783        & 2.5197       & 80.8860       \\ \hline
			\textbf{39}                         & 38.6120       & 0.9006        & 0.9110           & 37.2448       & 34.3887       & 31.3661      & 0.7586       & 0.5013       & 06018         & 2.287        & 78.2446       \\ \hline
			\textbf{43}                         & 35.1057       & 0.9002        & 0.9120           & 34.8161       & 31.8227       & 28.1837      & 0.6924       & 0.4506       & 0.5197        & 1.8260       & 76.1057       \\ \hline
			\multicolumn{12}{c}{(d)} \\ \hline
			\multirow{2}{*}{\textbf{QP levels}} & \multicolumn{11}{c|}{\textbf{FR-VQA performance metrics (Watersplashing)}}                                                                                                   \\ \cline{2-12} 
			& \textbf{PSNR} & \textbf{SSIM} & \textbf{MS-SSIM} & \textbf{VSNR} & \textbf{WSNR} & \textbf{NQM} & \textbf{UQI} & \textbf{VIF} & \textbf{VIFP} & \textbf{IFC} & \textbf{VMAF} \\ \hline
			\textbf{27}                         & 47.6243       & 0.9311        & 0.9470           & 44.1286       & 41.8786       & 40.8857      & 0.7982       & 0.7123       & 0.8344        & 3.3266       & 85.1581       \\ \hline
			\textbf{31}                         & 44.3516       & 0.9240        & 0.9547           & 42.4581       & 39.5503       & 38.1883      & 0.7902       & 0.6317       & 0.7365        & 2.9903       & 83.4137       \\ \hline
			\textbf{35}                         & 40.1624       & 0.9157        & 0.9310           & 38.0896       & 35.0068       & 34.0058      & 0.7663       & 0.5307       & 0.6167        & 2.4002       & 80.2735       \\ \hline
			\textbf{39}                         & 37.7736       & 0.9041        & 0.9082           & 36.5031       & 33.1732       & 30.2585      & 0.7381       & 0.4907       & 0.5984        & 2.1846       & 79.5284       \\ \hline
			\textbf{43}                         & 34.6053       & 0.8901        & 0.9257           & 34.2448       & 31.5025       & 28.0013      & 0.6603       & 0.4155       & 0.4794        & 1.7005       & 75.4002       \\ \hline
			\multicolumn{12}{c}{(e)} \\ 	
		\end{tabular}
	}
\end{table*} 
\par 
Table. IV shows the performance of HEVC using FR-VQA metrics at 30fps for different EVs at different QP levels. It can be seen that for each EVSs, as the QP level increases, the FR-VQA metrics vary correspondingly just like the subjective quality assessment measurements as discussed in section VI(A). Table IV also shows that the impact of increasing the frame rate results in an increase in the visual quality \cite{mackin2015study, usman2018novel}. For QP ranging within 27-31, all EVSs show higher values of PSNR, SSIM, WSNR, UQI, VIF, and VMAF being reflected as $\it{Excellent}$ quality. Similarly, as seen in Tables. IV(b) and IV(e), the FR-VQA metrics within QP ranging 39-43 show lower values for FR-VQA metrics being reflected as $\it{Poor}$ quality. The QP ranging within 39-43 in Tables. IV(a), IV(c), and IV(d) shows higher values reflected as a $\it{Good}$ visual quality. These results are highly correlated with the subjective measurements as shown in Fig. 3(b). Table. IV also shows that the network with low-bit rate requirements at 30fps EVs indicates acceptable quality using HEVC ranges within 31-35, based on both subjective and objective models.
\begin{table*}[]
	\caption{FR-VQM objective quality models corresponding each clip at 15fps for VP9.}
	\resizebox{18.00cm}{!}{
		\renewcommand{\arraystretch}{1.22}
		\centering
		\begin{tabular}{|c|c|c|c|c|c|c|c|c|c|c|c|}
			\hline
			\multirow{2}{*}{\textbf{QP levels}} & \multicolumn{11}{c|}{\textbf{FR-VQA performance metrics for EVs Books}}                                                                                                            \\ \cline{2-12} 
			& \textbf{PSNR} & \textbf{SSIM} & \textbf{MS-SSIM} & \textbf{VSNR} & \textbf{WSNR} & \textbf{NQM} & \textbf{UQI} & \textbf{VIF} & \textbf{VIFP} & \textbf{IFC} & \textbf{VMAF} \\ \hline
			\textbf{27}                         & \textbf{38.1824}       & \textbf{0.9205}        & 0.9257           & 34.0822       & \textbf{32.0057}       & 31.6052      & \textbf{0.7402}       & \textbf{0.4153}       & 0.5261        & 2.2001       & \textbf{83.2201}       \\ \hline
			\textbf{31}                         & \textbf{35.1801}       & \textbf{0.9180}        & 0.9105           & 31.0264       & \textbf{30.0010}       & 28.4406      & \textbf{0.6811}       & \textbf{0.3250}       & 0.4241        & 1.5807       & \textbf{80.1158}       \\ \hline
			\textbf{35}                         & 31.7183       & 0.8504        & 0.8620           & 30.7628       & 28.1836       & 26.3881      & \textbf{0.6108}       & 0.2607       & 0.3694        & 1.1662       & 75.7514       \\ \hline
			\textbf{39}                         & 28.0241       & 0.8102        & 0.8682           & 27.1572       & 26.3175       & 24.2357      & 0.5160       & 0.2304       & 0.3125        & 0.9858       & 75.0375       \\ \hline
			\textbf{43}                         & 25.2260       & 0.7601        & 0.8814           & 25.1602       & 24.6763       & 19.8164      & 0.4382       & 0.2081       & 0.3011        & 0.7022       & 72.8861       \\ \hline
			\multicolumn{12}{c}{(a)} \\ \hline		
			
			\multirow{2}{*}{\textbf{QP levels}} & \multicolumn{11}{c|}{\textbf{FR-VQA performance metrics for Evs Catchtrack}}                                                                                                       \\ \cline{2-12} 
			& \textbf{PSNR} & \textbf{SSIM} & \textbf{MS-SSIM} & \textbf{VSNR} & \textbf{WSNR} & \textbf{NQM} & \textbf{UQI} & \textbf{VIF} & \textbf{VIFP} & \textbf{IFC} & \textbf{VMAF} \\ \hline
			\textbf{27}                         & \textbf{37.3307}       & \textbf{0.8713}        & 0.9118           & 31.8647       & \textbf{30.6641}       & 30.7558      & \textbf{0.7016}       & \textbf{0.4002}       & 0.5101        & 1.9833       & \textbf{80.5734}       \\ \hline
			\textbf{31}                         & \textbf{34.2264}       & \textbf{0.8285}        & 0.8810           & 30.3527       & 28.8861       & 27.6138      & \textbf{0.6235}       & 0.3007       & 0.4006        & 1.3015       & \textbf{76.8803}       \\ \hline
			\textbf{35}                         & 29.5508       & \textbf{0.8580}        & 0.8741           & 30.0257       & 27.7506       & 26.0628      & \textbf{0.5892}       & 0.2408       & 0.3458        & 1.1078       & 75.0026       \\ \hline
			\textbf{39}                         & 27.6437       & 0.7791        & 0.8814           & 26.8581       & 25.8721       & 23.2606      & 0.4702       & 0.2287       & 0.3018        & 1.1057       & \textbf{75.1830}       \\ \hline
			\textbf{43}                         & 24.6247       & 0.7028        & 0.8930           & 23.8804       & 24.0176       & 18.6081      & 0.4206       & 0.1986       & 0.2884        & 0.6567       & 71.5033       \\ \hline		
			\multicolumn{12}{c}{(b)} \\ \hline		
			
			\multirow{2}{*}{\textbf{QP levels}} & \multicolumn{11}{c|}{\textbf{FR-VQA performance metrics for EVs Hamster}}                                                                                                          \\ \cline{2-12} 
			& \textbf{PSNR} & \textbf{SSIM} & \textbf{MS-SSIM} & \textbf{VSNR} & \textbf{WSNR} & \textbf{NQM} & \textbf{UQI} & \textbf{VIF} & \textbf{VIFP} & \textbf{IFC} & \textbf{VMAF} \\ \hline
			\textbf{27}                         & \textbf{38.3620}       & \textbf{0.9281}        & 0.9204           & 33.2016       & \textbf{31.4062}       & 31.0563      & \textbf{0.7463}       & \textbf{0.4251}       & 0.5206        & 2.2115       & \textbf{82.8862}       \\ \hline
			\textbf{31}                         & \textbf{35.0301}       & \textbf{0.8611}        & 0.8901           & 30.8842       & 29.6516       & 28.3722      & \textbf{0.6783}       & 0.3201       & 0.4182        & 1.5724       & \textbf{80.1757}       \\ \hline
			\textbf{35}                         & 31.1660       & \textbf{0.8427}        & 0.8904           & 30.2146       & 28.1042       & 26.1005      & \textbf{0.6004}       & 0.2513       & 0.3660        & 1.1531       & \textbf{76.6813}       \\ \hline
			\textbf{39}                         & 28.0131       & 0.8001        & 0.8906           & 27.2884       & 26.5086       & 24.2218      & 0.5001       & 0.2366       & 0.3427        & 1.1536       & 74.8862       \\ \hline
			\textbf{43}                         & 25.8901       & 0.7650        & 0.9050           & 25.2738       & 24.7714       & 20.4837      & 0.4410       & 0.2125       & 0.3116        & 0.7134       & 72.4706       \\ \hline	
			\multicolumn{12}{c}{(c)} \\ \hline	
			\multirow{2}{*}{\textbf{QP levels}} & \multicolumn{11}{c|}{\textbf{FR-VQA performance metrics for EVs Sparkler}}                                                                                                         \\ \cline{2-12} 
			& \textbf{PSNR} & \textbf{SSIM} & \textbf{MS-SSIM} & \textbf{VSNR} & \textbf{WSNR} & \textbf{NQM} & \textbf{UQI} & \textbf{VIF} & \textbf{VIFP} & \textbf{IFC} & \textbf{VMAF} \\ \hline
			\textbf{27}                         & \textbf{37.8244}       & \textbf{0.8905}        & 0.9037           & 33.9047       & \textbf{31.1006}       & 30.8804      & \textbf{0.7175}       & \textbf{0.4168}       & 0.5118        & 0.2108       & \textbf{82.9016}       \\ \hline
			\textbf{31}                         & \textbf{35.6120}       & \textbf{0.9003}        & 0.8980           & 31.4061       & \textbf{30.2036}       & 28.8806      & \textbf{0.6907}       & \textbf{0.3316}       & 0.4207        & 1.6646       & \textbf{80.1840}       \\ \hline
			\textbf{35}                         & 31.7361       & \textbf{0.9001}        & 0.8630           & 30.7703       & 28.1426       & 26.4508      & \textbf{0.6117}       & 0.2688       & 0.3696        & 1.1697       & \textbf{76.2064}       \\ \hline
			\textbf{39}                         & 28.7220       & \textbf{0.8204}        & 0.8440           & 27.6833       & 26.8943       & 24.6084      & 0.5180       & 0.2418       & 0.3306        & 1.1602       & \textbf{75.1748}       \\ \hline
			\textbf{43}                         & 25.8830       & 0.7703        & 0.8311           & 25.7551       & 24.7806       & 20.3751      & 0.4496       & 0.2107       & 0.3101        & 0.7101       & 72.3050       \\ \hline
			\multicolumn{12}{c}{(d)} \\ \hline
			\multirow{2}{*}{\textbf{QP levels}} & \multicolumn{11}{c|}{\textbf{FR-VQA performance metrics for EVs Watersplashing}}                                                                                                   \\ \cline{2-12} 
			& \textbf{PSNR} & \textbf{SSIM} & \textbf{MS-SSIM} & \textbf{VSNR} & \textbf{WSNR} & \textbf{NQM} & \textbf{UQI} & \textbf{VIF} & \textbf{VIFP} & \textbf{IFC} & \textbf{VMAF} \\ \hline
			\textbf{27}                         & \textbf{37.0120}       & \textbf{0.8864}        & 0.9102           & 33.0615       & \textbf{30.2263}      & 30.3826      & \textbf{0.7001}       & \textbf{0.3937}       & 0.4884        & 0.1886       & \textbf{80.3370}       \\ \hline
			\textbf{31}                         & \textbf{34.1033}       & \textbf{0.8147}        & 0.8980           & 30.1537       & 29.2557       & 28.5571      & \textbf{0.6210}       & 0.3001       & 0.3884        & 1.3004       & \textbf{76.4061}       \\ \hline
			\textbf{35}                         & 29.0014       & \textbf{0.8206}        & 0.8507           & 29.6216       & 27.4013       & 25.8735      & \textbf{0.5801}       & 0.2004       & 0.3184        & 1.0074       & \textbf{75.4081}       \\ \hline
			\textbf{39}                         & 27.4063       & 0.78060       & 0.8030           & 26.5502       & 25.6807       & 23.1061      & 0.4684       & 0.2137       & 0.2905        & 0.9805       & 74.6063       \\ \hline
			\textbf{43}                         & 25.0046       & 0.7007        & 0.8220           & 25.1527       & 24.6248       & 19.9007      & 0.4147       & 0.2003       & 0.3002        & 0.6897       & 71.8346       \\ \hline
			\multicolumn{12}{c}{(e)} 
		\end{tabular}
	}
\end{table*} 
\par 
Similarly, Table. V shows the performance of HEVC using FR-VQA metrics at 60fps for different EVs at different QP levels. It can be seen that for each EVSs, as the QP level increases, the FR-VQA metrics vary correspondingly, just like the subjective quality assessment measurements discussed in section VI(A). Table. V also shows the impact of increasing the frame rate on increasing visual quality \cite{mackin2015study, mackin2017investigating, usman2018novel}. As shown in Fig. 3(c), the subjective measurements for all EVSs at QP ranging within 39-43 reflects $\it{Fair}$ visual quality, Table. V confirms, in a correlated fashion, higher values of FR-VQA metrics within the same QP ranges. As observed in Tables. V(a), V(c), and V(d), the QP values ranging within 35-39 show higher values for FR-VQA metrics being reflecting as a $\it{Good}$ visual quality, while for the QP values ranging within 27-35 reflect an $\it{Excellent}$ visual quality. These results are highly correlated with the subjective measurements shown in Fig. 3(c). Note that the perceptual visual quality in terms of DMOS and FR objective metrics increases with the frame rate. Table. V also shows that for a network with low bit-rate requirements at 60fps EVSs, an acceptable quality is achieved using HEVC ranging within 27- 39 based, on both subjective and objective models. 
\subsubsection{FR Objective VQA Results for VP9}
Considering the 11 state-of-the-art FR-objective VQA metrics discussed in section V, including the recently adopted VAMF trained on NETFLIX's videos \cite{netflix2019video}, we attempt to determine the relation between subjective and objective measurements for benchmarking. Detailed performance of the opted FR-VQA at all three frame rates using VP9 at 15fps, 30fps, and 60fps is given in Table. VI, Table. VII, and Table. VIII, respectively.

\begin{table*}[]
	\caption{FR-VQM objective quality models corresponding each clip at 30fps for VP9.}
	\resizebox{18.00cm}{!}{
		\renewcommand{\arraystretch}{1.22}
		\centering
		\begin{tabular}{|c|c|c|c|c|c|c|c|c|c|c|c|}
			\hline
			\multirow{2}{*}{\textbf{QP levels}} & \multicolumn{11}{c|}{\textbf{FR-VQA performance metrics for EVs Books}}                                                                                                            \\ \cline{2-12} 
			& \textbf{PSNR} & \textbf{SSIM} & \textbf{MS-SSIM} & \textbf{VSNR} & \textbf{WSNR} & \textbf{NQM} & \textbf{UQI} & \textbf{VIF} & \textbf{VIFP} & \textbf{IFC} & \textbf{VMAF} \\ \hline
			\textbf{27}                         & \textbf{43.4010}       & \textbf{0.9305}        & 0.9386           & 40.1246       & \textbf{38.2066}       & 35.5538      & \textbf{0.7724}       & \textbf{0.5126}       & 0.6088        & 2.5006       & \textbf{84.2206}       \\ \hline
			\textbf{31}                         & \textbf{40.0214}       & \textbf{0.9224}        & 0.9246           & 36.8461       & \textbf{34.2608}       & 34.6126      & \textbf{0.7430}       & \textbf{0.5003}       & 0.5927        & 2.3156       & \textbf{81.1852}       \\ \hline
			\textbf{35}                         & \textbf{36.2203}       & \textbf{0.8901}        & 0.9004           & 33.7752       & 29.8906       & 30.6187      & \textbf{0.6807}       & \textbf{0.3912}       & 0.4805        & 1.5583       & \textbf{78.2641}       \\ \hline
			\textbf{39}                         & \textbf{34.1257}       & \textbf{0.8480}        & 0.8601           & 30.5574       & 27.2833       & 25.2833      & 0.6447       & \textbf{0.3617}       & 0.4406        & 1.3744       & \textbf{76.6224}       \\ \hline
			\textbf{43}                         & 29.7216       & 0.8130        & 0.8861           & 28.2684       & 26.3442       & 22.8482      & 0.5531       & 0.3007       & 0.3814        & 1.1835       & 73.2663       \\ \hline
			\multicolumn{12}{c}{(a)} \\ \hline				
			\multirow{2}{*}{\textbf{QP levels}} & \multicolumn{11}{c|}{\textbf{FR-VQA performance metrics for EVs Catchtrack}}                                                                                                       \\ \cline{2-12} 
			& \textbf{PSNR} & \textbf{SSIM} & \textbf{MS-SSIM} & \textbf{VSNR} & \textbf{WSNR} & \textbf{NQM} & \textbf{UQI} & \textbf{VIF} & \textbf{VIFP} & \textbf{IFC} & \textbf{VMAF} \\ \hline
			\textbf{27}                         & \textbf{43.0286}       & \textbf{0.9288}        & 0.9304           & 40.0051       & \textbf{38.0617}       & 35.2361      & \textbf{0.7611}       & \textbf{0.5003}       & 0.6001        & 2.4883       & \textbf{82.4668}       \\ \hline
			\textbf{31}                         & \textbf{39.8806}       & \textbf{0.8947}        & 0.9041           & 35.2873       & \textbf{33.8791}       & 34.4001      & \textbf{0.7262}       & \textbf{0.4257}       & 0.5361        & 2.0668       & \textbf{81.3360}       \\ \hline
			\textbf{35}                         & \textbf{35.8478}       & \textbf{0.8620}        & 0.8726           & 32.6884       & 28.6873       & 29.4037      & \textbf{0.6724}       & \textbf{0.3801}       & 0.4711        & 1.4907       & \textbf{78.1204}       \\ \hline
			\textbf{39}                         & \textbf{33.2406}       & \textbf{0.8503}        & 0.8520           & 29.7617       & 26.8878       & 24.7684      & \textbf{0.6210}       & 0.3125       & 0.4072        & 1.2364       & 74.8871       \\ \hline
			\textbf{43}                         & 27.0245       & 0.7664        & 0.8430           & 27.5591       & 25.6268       & 21.1264      & 0.4916       & 0.2657       & 0.3351        & 1.1261       & 72.8910       \\ \hline		
			\multicolumn{12}{c}{(b)} \\ \hline		
			\multirow{2}{*}{\textbf{QP levels}} & \multicolumn{11}{c|}{\textbf{FR-VQA performance metrics for EVs Hamster}}                                                                                                          \\ \cline{2-12} 
			& \textbf{PSNR} & \textbf{SSIM} & \textbf{MS-SSIM} & \textbf{VSNR} & \textbf{WSNR} & \textbf{NQM} & \textbf{UQI} & \textbf{VIF} & \textbf{VIFP} & \textbf{IFC} & \textbf{VMAF} \\ \hline
			\textbf{27}                         & \textbf{43.8722}       & \textbf{0.9311}        & 0.9362           & 40.5287       & \textbf{38.4882}       & 35.8892      & \textbf{0.7788}       & \textbf{0.5481 }      & 0.6135        & 2.5128       & \textbf{84.0053}       \\ \hline
			\textbf{31}                         & \textbf{40.6108}       & \textbf{0.9280}        & 0.9330           & 36.8812       & \textbf{34.2682}       & 34.8863      & \textbf{0.7486}       & \textbf{0.4838 }      & 0.5736        & 2.1288       & \textbf{81.7724}       \\ \hline
			\textbf{35}                         & \textbf{36.6021}       & \textbf{0.8926 }       & 0.9010           & 33.8837       & 29.9064       & 30.7728      & \textbf{0.6935}       & \textbf{0.3975}       & 0.4903        & 1.5746       & \textbf{79.0030}      \\ \hline
			\textbf{39}                         & \textbf{34.2613}       & \textbf{0.8527}        & 0.8615           & 30.5864       & 27.8643       & 25.3168      & \textbf{0.6488}       & \textbf{0.3688}       & 0.4484        & 1.3897       & \textbf{76.8914}       \\ \hline
			\textbf{43}                         & 29.7615       & \textbf{0.8286}        & 0.8893           & 28.4583       & 26.4677       & 22.0188      & \textbf{0.5607}       & 0.3124       & 0.3881        & 1.1984       & 73.2894       \\ \hline		
			\multicolumn{12}{c}{(c)} \\ \hline
			
			\multirow{2}{*}{\textbf{QP levels}} & \multicolumn{11}{c|}{\textbf{FR-VQA performance metrics for EVs Sparkler}}                                                                                                         \\ \cline{2-12} 
			& \textbf{PSNR} & \textbf{SSIM} & \textbf{MS-SSIM} & \textbf{VSNR} & \textbf{WSNR} & \textbf{NQM} & \textbf{UQI} & \textbf{VIF} & \textbf{VIFP} & \textbf{IFC} & \textbf{VMAF} \\ \hline
			\textbf{27}                         & \textbf{43.9005}       & \textbf{0.9355}        & 0.9201           & 40.5886       & \textbf{35.5571}       & 36.0151      & \textbf{0.7816}       & \textbf{0.5582}       & 0.6204        & 2.5176       & \textbf{82.5570}       \\ \hline
			\textbf{31}                         & \textbf{40.5584}       & \textbf{0.9284}        & 0.9362           & 36.8786       & \textbf{34.2503}       & 34.5579      & \textbf{0.7488}       & \textbf{0.4702}       & 0.5602        & 2.0157       & \textbf{81.6405}       \\ \hline
			\textbf{35}                         & \textbf{36.4463}       & \textbf{0.8811}        & 0.8947           & 33.7873       & 29.8961       & 30.5013      & \textbf{0.6911}       & \textbf{0.3941}       & 0.4841        & 1.5620       & \textbf{79.0153}       \\ \hline
			\textbf{39}                         & \textbf{34.0370}       & \textbf{0.8511}        & 0.8577           & 30.2594       & 27.4906       & 25.2001      & \textbf{0.6173}       & 0.3407       & 0.4157        & 1.3201       & \textbf{76.7033}       \\ \hline
			\textbf{43}                         & 28.8341       & 0.8001        & 0.8970           & 28.1512       & 26.2003       & 21.9004      & 0.5244       & 0.3001       & 0.3706        & 1.1662       & 73.9430       \\ \hline
			\multicolumn{12}{c}{(d)} \\ \hline
			\multirow{2}{*}{\textbf{QP levels}} & \multicolumn{11}{c|}{\textbf{FR-VQA performance metrics for EVs Watersplashing}}                                                                                                   \\ \cline{2-12} 
			& \textbf{PSNR} & \textbf{SSIM} & \textbf{MS-SSIM} & \textbf{VSNR} & \textbf{WSNR} & \textbf{NQM} & \textbf{UQI} & \textbf{VIF} & \textbf{VIFP} & \textbf{IFC} & \textbf{VMAF} \\ \hline
			\textbf{27}                         & \textbf{43.0167}       & \textbf{0.8984}        & 0.9007           & 40.0016       & \textbf{37.8864}       & 35.1283      & \textbf{0.7603}       & \textbf{0.5007}       & 0.6115        & 2.4985       & \textbf{81.9004}       \\ \hline
			\textbf{31}                         & \textbf{39.4833}       & \textbf{0.8836}        & 0.9006           & 35.0159       & \textbf{33.0067}       & 33.8991      & \textbf{0.7116}      & \textbf{0.4105}       & 0.5183        & 1.9804       & \textbf{80.6240}       \\ \hline
			\textbf{35}                         & \textbf{35.6801}       & \textbf{0.8581}         & 0.8643           & 32.4006       & 28.2684       & 29.2691      & \textbf{0.6880}       & \textbf{0.3707}       & 0.4537        & 1.4016       & \textbf{78.4026}       \\ \hline
			\textbf{39}                         & \textbf{33.0058}       & \textbf{0.8246}        & 0.8511           & 29.5803       & 26.6833       & 24.5086      & \textbf{0.6178}       & 0.3104       & 0.3910        & 1.2118       & 74.0461       \\ \hline
			\textbf{43}                         & 26.9024       & 0.7154        & 0.8066           & 27.0574       & 28.0886       & 21.0667      & 0.5168       & 0.2508       & 0.3288        & 0.9164       & 71.9970       \\ \hline
			\multicolumn{12}{c}{(e)} \\ 	
		\end{tabular}
	}
\end{table*}
\par 
Table. VI shows the performance of VP9 in terms of FR-VQA metrics for 15fps EVSs at different QP levels. A uniform variation can be observed for the FR-VQA metrics as the QP level varies correspondingly, as shown in Fig. 4, during the subjective quality assessment discussed in section VI(A) due to compression. VP9 which is customized for video resolutions beyond 1080p and lossless compression, is 20\% less efficient than HEVC and requires two times the bit-rate to reach the same quality that is achieved by HEVC. H.265/HEVC still outperforms in terms of visual quality in terms for FR-VQA metrics at the cost of encoding time \cite{barman2017h}. For 15fps, the FR-VQA metrics such as PSNR, SSIM, WSNR, UQI, VIF, and VMAF, show better performance, as shown in Table. VI. For QP ranging within 27-31, all EVSs show higher values of PSNR, SSIM, WSNR, UQI, VIF, and VMAF reflected as $\it{Excellent}$ quality. However, the FR-VQA metrics in Tables. VI(b) and VI(e) within the QP range of 39-43 show a $\it{Poor}$ quality, while in the same QP range, a $\it{Fair}$  quality is observed, as shown in Tables. VI(a), VI(c), and VI(d). Furthermore, the FR-VQA metric performance using VP9 within QP range 31-35 in Tables. VI(a), VI(c), and VI(d) reflects $\it{Good}$ quality, while in the same QP range, the reflected quality in Tables. VI(b) and VI(e) is $\it{Fair}$. The behavior of FR-VQA metrics is highly similar to that observed during the subjective quality assessment ,as shown in Fig. 4(a). 
\begin{table*}[]
	\caption{FR-VQM objective quality models corresponding each clip at 60fps for VP9.}
	\resizebox{18.00cm}{!}{
		\renewcommand{\arraystretch}{1.22}
		\centering
		\begin{tabular}{|c|c|c|c|c|c|c|c|c|c|c|c|}
			\hline
			\multirow{2}{*}{\textbf{QP levels}} & \multicolumn{11}{c|}{\textbf{FR-VQA performance metrics for EVs Books}}                                                                                                            \\ \cline{2-12} 
			& \textbf{PSNR} & \textbf{SSIM} & \textbf{MS-SSIM} & \textbf{VSNR} & \textbf{WSNR} & \textbf{NQM} & \textbf{UQI} & \textbf{VIF} & \textbf{VIFP} & \textbf{IFC} & \textbf{VMAF} \\ \hline
			\textbf{27}                         & 46.6423       & 0.9506        & 0.9601           & 44.1006       & 41.6228       & 39.7655      & 0.8203       & 0.6882       & 0.7891        & 2.9816       & 85.9241       \\ \hline
			\textbf{31}                         & 42.8046       & 0.9310        & 0.9404           & 39.6084       & 36.7624       & 36.0289      & 0.7682       & 0.6013       & 0.7005        & 2.7641       & 82.1617       \\ \hline
			\textbf{35}                         & 38.6814       & 0.9101        & 0.9124           & 35.9075       & 32.8613       & 33.0256      & 0.7201       & 0.4904       & 0.5881        & 2.1682       & 81.3518       \\ \hline
			\textbf{39}                         & 36.6120       & 0.8481        & 0.8788           & 32.2678       & 29.8062       & 27.8162      & 0.6742       & 0.4462       & 0.5527        & 1.7942       & 77.8615       \\ \hline
			\textbf{43}                         & 32.2861       & 0.8184        & 0.8553           & 31.1673       & 29.0025       & 24.6315      & 0.5916       & 0.4025       & 0.4651        & 1.5681       & 75.0836       \\ \hline
			\multicolumn{12}{c}{(a)} \\ \hline		
			\multirow{2}{*}{\textbf{QP levels}} & \multicolumn{11}{c|}{\textbf{FR-VQA performance metrics for EVs Catchtrack}}                                                                                                       \\ \cline{2-12} 
			& \textbf{PSNR} & \textbf{SSIM} & \textbf{MS-SSIM} & \textbf{VSNR} & \textbf{WSNR} & \textbf{NQM} & \textbf{UQI} & \textbf{VIF} & \textbf{VIFP} & \textbf{IFC} & \textbf{VMAF} \\ \hline
			\textbf{27}                         & 45.8467       & 0.9442        & 0.9422           & 43.2576       & 40.2881       & 38.3617      & 0.7986       & 0.6407       & 0.7357        & 2.7977       & 83.2750       \\ \hline
			\textbf{31}                         & 41.0057       & 0.9241        & 0.9136           & 38.8916       & 35.8225       & 35.8905      & 0.7516       & 0.5568       & 0.6421        & 2.4915       & 80.3581       \\ \hline
			\textbf{35}                         & 37.2003       & 0.8952        & 0.9034           & 35.0726       & 32.1288       & 32.7257      & 0.6915       & 0.4716       & 0.5724        & 2.0226       & 80.5481       \\ \hline
			\textbf{39}                         & 35.0362       & 0.8203        & 0.8511           & 32.1613       & 29.6031       & 27.0561      & 0.6465       & 0.4301       & 0.5386        & 1.7274       & 77.3546       \\ \hline
			\textbf{43}                         & 31.3581       & 0.7752        & 0.8136           & 30.0689       & 28.1277       & 23.6273      & 0.5863       & 0.3984       & 0.4588        & 1.5066       & 74.5811       \\ \hline		
			
			\multicolumn{12}{c}{(b)} \\ \hline	
			\multirow{2}{*}{\textbf{QP levels}} & \multicolumn{11}{c|}{\textbf{FR-VQA performance metrics for EVs Hamster}}                                                                                                          \\ \cline{2-12} 
			& \textbf{PSNR} & \textbf{SSIM} & \textbf{MS-SSIM} & \textbf{VSNR} & \textbf{WSNR} & \textbf{NQM} & \textbf{UQI} & \textbf{VIF} & \textbf{VIFP} & \textbf{IFC} & \textbf{VMAF} \\ \hline
			\textbf{27}                         & 46.8640       & 0.9517        & 0.9634           & 43.2681       & 40.2906       & 38.3884      & 0.8104       & 0.7011       & 0.8015        & 3.1253       & 85.9330       \\ \hline
			\textbf{31}                         & 42.8824       & 0.9336        & 0.9227           & 40.1258       & 37.5508       & 36.6134      & 0.7688       & 0.6077       & 0.7101        & 2.7881       & 82.3350       \\ \hline
			\textbf{35}                         & 38.8836       & 0.9155        & 0.9220           & 36.1566       & 33.2664       & 32.8806      & 0.7284       & 0.5005       & 0.5988        & 2.1894       & 81.8930       \\ \hline
			\textbf{39}                         & 36.8812       & 0.8584        & 0.9013           & 33.8467       & 30.5335       & 28.1683      & 0.6803       & 0.4583       & 0.5611        & 1.8015       & 78.0150       \\ \hline
			\textbf{43}                         & 32.6603       & 0.8210        & 0.8904           & 31.8066       & 29.5524       & 24.8904      & 0.6107       & 0.4088       & 0.4697        & 1.8914       & 75.2566       \\ \hline				
			\multicolumn{12}{c}{(c)} \\ \hline	
			\multirow{2}{*}{\textbf{QP levels}} & \multicolumn{11}{c|}{\textbf{FR-VQA performance metrics for EVs Sparkler}}                                                                                                         \\ \cline{2-12} 
			& \textbf{PSNR} & \textbf{SSIM} & \textbf{MS-SSIM} & \textbf{VSNR} & \textbf{WSNR} & \textbf{NQM} & \textbf{UQI} & \textbf{VIF} & \textbf{VIFP} & \textbf{IFC} & \textbf{VMAF} \\ \hline
			\textbf{27}                         & 46.7301       & 0.9500        & 0.9510           & 44.1806       & 41.8830       & 38.4806      & 0.8001       & 0.6914       & 0.7897        & 3.0158       & 84.8864       \\ \hline
			\textbf{31}                         & 42.8361       & 0.9315        & 0.9384           & 40.0581       & 37.2643       & 36.5118      & 0.7690       & 0.5921       & 0.6824        & 2.6924       & 82.3010       \\ \hline
			\textbf{35}                         & 38.7215       & 0.9087        & 0.8870           & 36.1483       & 33.2107       & 33.5015      & 0.7260       & 0.4922       & 0.5913        & 2.1896       & 81.8806       \\ \hline
			\textbf{39}                         & 36.8764       & 0.8561        & 0.8840           & 33.9041       & 30.8133       & 28.1062      & 0.6812       & 0.4511       & 0.5586        & 1.8158       & 78.0080       \\ \hline
			\textbf{43}                         & 32.8417       & 0.8218        & 0.8910           & 31.9057       & 29.6006       & 25.0068      & 0.6180       & 0.4110       & 0.4803        & 1.6603       & 75.2606       \\ \hline
			\multicolumn{12}{c}{(d)} \\ \hline		
			\multirow{2}{*}{\textbf{QP levels}} & \multicolumn{11}{c|}{\textbf{FR-VQA performance metrics for EVs Watersplashing}}                                                                                                   \\ \cline{2-12} 
			& \textbf{PSNR} & \textbf{SSIM} & \textbf{MS-SSIM} & \textbf{VSNR} & \textbf{WSNR} & \textbf{NQM} & \textbf{UQI} & \textbf{VIF} & \textbf{VIFP} & \textbf{IFC} & \textbf{VMAF} \\ \hline
			\textbf{27}                         & 45.3352       & 0.9473        & 0.9246           & 43.0174       & 40.1007       & 38.0028      & 0.7880       & 0.6342       & 0.7261        & 2.8022       & 82.8162       \\ \hline
			\textbf{31}                         & 40.9023       & 0.9103        & 0.9063           & 38.0573       & 35.1681       & 35.0027      & 0.7471       & 0.5284       & 0.6236        & 2.5057       & 79.8862       \\ \hline
			\textbf{35}                         & 37.4007       & 0.8981        & 0.9006           & 35.2287       & 32.3891       & 32.8201      & 0.7006       & 0.4738       & 0.5775        & 2.0551       & 80.5680       \\ \hline
			\textbf{39}                         & 34.7283       & 0.8298        & 0.8362           & 32.0037       & 29.1247       & 27.0286      & 0.6304       & 0.4275       & 0.5307        & 1.7992       & 77.1006       \\ \hline
			\textbf{43}                         & 31.4050       & 0.7603        & 0.8107           & 30.1863       & 28.2683       & 24.8735      & 0.5902       & 0.3807       & 0.4513        & 1.4284       & 74.7008       \\ \hline
			\multicolumn{12}{c}{(e)}		
		\end{tabular}
	}
\end{table*}
\par 
Table. VII shows the performance of VP9 in terms of FR-VQA metrics for 30fps EVSs at different QP levels. It can be observed in Table. VII that, for each EVSs as QP level increases, the performance of FR-VQA metrics varies correspondingly, similar to the subjective quality measurements, as discussed in section VI(A) and shown in Fig. 4(b). Table. VII also proves that as the frame rate increases, the visual quality in terms of FR-VQA such as PSNR, SSIM, WSNR, UQI, VIF, and VMAF metrics, is better than that presented in Table. VI. For QP ranging within 27-31, all EVSs show higher values of PSNR, SSIM, WSNR, UQI, VIF, and VMAF being reflected as $\it{Excellent}$ quality. However, the FR-VQA metrics within the QP range of 39-43 show a $\it{Fair}$ quality for all EVSs, as shown in Table. VII. Similarly, the FR-VQA metric performance obtained using VP9 within the QP range 31-35 in Tables. VI(a), VI(c), and VI(d) reflects $\it{Good}$ quality, while within the same QP range, the reflected quality in Tables. VI (b) and VI(e) is $\it{Fair}$. The behavior of FR-VQA metrics is highly similar to that observed during the subjective quality assessment, as shown in Fig. 4(b). Tables. VI and VII, clearly show that the impact of frame rate, i.e., from 15fps to 30fps has medial significance on visual quality when VP9 is used as the encoder. The significance on the visual quality is vivid in the case of H.265/HEVC for all EVSs at each QP level. 

\begin{table*}[t]
	\renewcommand{\arraystretch}{1.7}
	\caption{Correlation coefficient scores (PLCC and SROCC) between FR-VQA metrics and DMOS for H.265/HEVC at (15fps, 30fps, and 60fps).}
	\resizebox{18.00cm}{!}{
		\begin{tabular}{|c|c|c|c|c|c|c|c|c|c|c|c|c|}
			\hline
			\multirow{2}{*}{\textbf{Frame rates}} & \textbf{}                        & \multicolumn{11}{c|}{\textbf{FR-VQA performance metrics for H.265/HEVC}}                                                                                                                     \\ \cline{2-13} 
			& \textbf{Correlation Coefficient} & \textbf{PSNR}   & \textbf{SSIM}   & \textbf{MS-SSIM} & \textbf{VSNR}   & \textbf{WSNR}   & \textbf{NQM} & \textbf{UQI}    & \textbf{VIF}    & \textbf{VIFP} & \textbf{IFC} & \textbf{VMAF}   \\ \hline
			\multirow{2}{*}{\textbf{15fps}}       & \textbf{PLCC}                    & \textbf{0.8783} & 0.8157          & 0.8613           & 0.8411          & 0.8625          & 0.8048       & 0.8627          & \textbf{0.8814} & 0.8611        & 0.8642       & \textbf{0.8944} \\ \cline{2-13} 
			& \textbf{SROCC}                   & \textbf{0.8766} & 0.8354          & 0.8691           & 0.8533          & 0.8524          & 0.8157       & 0.8705          & \textbf{0.8892} & 0.8627        & 0.8714       & \textbf{0.9001} \\ \hline
			\multirow{2}{*}{\textbf{30fps}}       & \textbf{PLCC}                    & \textbf{0.9136} & 0.8504          & 0.8577           & 0.8611          & 0.8588          & 0.8502       & \textbf{0.8664} & \textbf{0.8984} & 0.8542        & 0.8684       & \textbf{0.9227} \\ \cline{2-13} 
			& \textbf{SROCC}                   & \textbf{0.9274} & 0.8661          & \textbf{0.8986}  & 0.8692          & \textbf{0.8944} & 0.8655       & 0.8648          & \textbf{0.9207} & 0.8897        & 0.8725       & \textbf{0.9304} \\ \hline
			\multirow{2}{*}{\textbf{60fps}}       & \textbf{PLCC}                    & \textbf{0.9481} & \textbf{0.8916} & \textbf{0.8955}  & \textbf{0.8891} & 0.9007          & 0.8922       & 0.9131          & \textbf{0.9486} & 0.8915        & 0.8841       & \textbf{0.9506} \\ \cline{2-13} 
			& \textbf{SROCC}                   & \textbf{0.9584} & \textbf{0.9117} & 0.9014           & 0.9002          & 0.9105          & 0.9011       & \textbf{0.9277} & \textbf{0.9566} & 0.9103        & 0.9001       & \textbf{0.9608} \\ \hline
		\end{tabular}
	}
\end{table*}

\begin{table*}[t]
	\renewcommand{\arraystretch}{1.7}
	\caption{Correlation coefficient scores (PLCC and SROCC) between FR-VQA metrics and DMOS for VP9 at (15fps, 30fps, and 60fps).}
	\resizebox{18.00cm}{!}{
	\begin{tabular}{|c|c|c|c|c|c|c|c|c|c|c|c|c|}
		\hline
		\multirow{2}{*}{\textbf{Frame rates}} & \textbf{}                        & \multicolumn{11}{c|}{\textbf{FR-VQA performance metrics for VP9}}                                                                                                                         \\ \cline{2-13} 
		& \textbf{Correlation Coefficient} & \textbf{PSNR}   & \textbf{SSIM}   & \textbf{MS-SSIM} & \textbf{VSNR}   & \textbf{WSNR}   & \textbf{NQM} & \textbf{UQI} & \textbf{VIF}    & \textbf{VIFP} & \textbf{IFC} & \textbf{VMAF}   \\ \hline
		\multirow{2}{*}{\textbf{15fps}}       & \textbf{PLCC}                    & \textbf{0.8622} & 0.8277          & 0.8587           & \textbf{0.8814} & \textbf{0.8801} & 0.8244       & 0.8381       & 0.8601          & 0.8413        & 0.8504       & \textbf{0.8875} \\ \cline{2-13} 
		& \textbf{SROCC}                   & \textbf{0.8814} & 0.8401          & 0.8602           & \textbf{0.8861} & \textbf{0.8847} & 0.8227       & 0.8784       & 0.8691          & 0.8627        & 0.8714       & \textbf{0.8994} \\ \hline
		\multirow{2}{*}{\textbf{30fps}}       & \textbf{PLCC}                    & \textbf{0.9253} & 0.8782          & 0.8771           & \textbf{0.9103} & \textbf{0.9117} & 0.8681       & 0.8822       & \textbf{0.8991} & 0.8624        & 0.8657       & \textbf{0.9304} \\ \cline{2-13} 
		& \textbf{SROCC}                   & \textbf{0.9288} & 0.8815          & 0.8853           & \textbf{0.9225} & \textbf{0.9264} & 0.8655       & 0.8648       & \textbf{0.9007} & 0.8877        & 0.8815       & \textbf{0.9355} \\ \hline
		\multirow{2}{*}{\textbf{60fps}}       & \textbf{PLCC}                    & \textbf{0.9381} & \textbf{0.9278} & 0.8905           & \textbf{0.9411} & \textbf{0.9376} & 0.9042       & 0.8984       & 0.8968          & 0.9022        & 0.8844       & \textbf{0.9491} \\ \cline{2-13} 
		& \textbf{SROCC}                   & \textbf{0.9603} & \textbf{0.9282} & 0.9007           & \textbf{0.9486} & \textbf{0.9503} & 0.9141       & 0.9077       & 0.9113          & 0.9087        & 0.9047       & \textbf{0.9683} \\ \hline
	\end{tabular}
}
\end{table*}
\par 
Similarly, Table. VIII shows the performance of VP9 using FR-VQA metrics at 60fps for different EVs at different QP levels. It can be seen that, for each EVSs, as the QP level increases, the FR-VQA metrics vary correspondingly, just like the subjective quality assessment measurements discussed in section VI(A). Table. VIII also shows the impact of increasing the frame rate on increasing visual quality \cite{mackin2015study, mackin2017investigating,usman2018novel}. As shown in Fig. 4(c), the subjective measurements for all EVSs at QP ranging within 39-43 reflect a $\it{Fair}$ visual quality. Table. VIII confirms, in a correlated fashion, higher values of FR-VQA metrics obtained within the same QP range. As observed in Tables. VIII(a), VIII(c), and VIII(d), QP values ranging within 35-39 show higher values of FR-VQA metrics being reflected as a $\it{Good}$ visual quality, while  QP values ranging within 27-35 reflect an $\it{Excellent}$ visual quality. These results are highly correlated with the subjective measurements as shown in Fig. 4(c). Note that the perceptual visual quality in terms of DMOS and FR objective metrics increases with the frame rate. Table. VIII also shows that for network with low bit-rate requirements at 60fps EVSs, the acceptable quality using VP9 ranges within 27- 39 based, on both subjective and objective models. Tables. VIII, clearly show that the impact of frame rate has better significance on visual quality when VP9 is used as the encoder at frame rate of 60fps. However, in comparison significance on the visual quality is vivid in the case of H.265/HEVC for all EVSs at each QP level. 
\subsection{Correlation between Subjective DMOS AND FR-VQMs Measurements:}
This section attempts to validate and shows the statistical evaluation of FR-VQM objective quality models listed in Table. (III-VIII). We have computed the Pearson's Linear Correlation Coefficient (PLCC) and the Spearman's Rank-Order Correlation Coefficient (SROCC) between the 11 state-of-the-art FR -VQA metrics and DMOS values using built-in MATLAB function. These statistical evaluation are conducted for both HEVC and VP9 and for all frame rates (15fps, 30fps, and 60fps) that are shown in Table. IX and Table. X, respectively. Noted, that the range of PLCC and SROCC is within [0-1], where close to 1 is preferable and indicated as a high correlation. 
\par 
Observed in Table. IX, it is infer that for all three frame rates (15fps, 30fps, and 60fps) employing H.265/HEVC shows a high a PLCC and SROCC values which is obtained between the DMOS and FR-VQA performance metrics. However, the FR-VQA metrics such as PSNR, WSNR, UQI, VIF, and VMAF shows a high PLCC and SROCC values for all three frame rates (15fps, 30fps, and 60fps). This can be interpreted that as, FHD contents encoded at different QP levels using H.265/HEVC, a high PLCC and SROCC values are generated reflecting that FR-VQA performance metrics such as PSNR, VIF, and VMAF are performing better and a recommended to be employed. 
\par 
Similarly, for all three frame rates (15fps, 30fps, and 60fps) employing VP9 shows a high a PLCC and SROCC values which is obtained between the DMOS and FR-VQA performance metrics. However, the FR-VQA metrics such as PSNR, VSNR, WSNR, VIF, and VMAF shows a high cc values for all three frame rates (15fps, 30fps, and 60fps). This can be interpreted that as, FHD contents encoded at different QP levels using VP9, a high PLCC and SROCC values are generated reflecting that FR-VQA performance metrics such as PSNR, VSNR, WSNR, and VMAF are performing better and a recommended to be employed. 
\section{Recommendations for Enhancing the Quality Estimation of FR-VQMs:}
All these aforementioned FR-VQMs are designed to estimate the spatial degradations of an image or a video sequence. Estimating the perceptual quality of compressed videos based on varying frame rates requires temporal degradation information. This information can be acquired by processing the compressed video sequences and extract certain features. These measurements or features should be included in the overall quality estimation of a video sequence.
\begin{itemize}
\item The first and most important measure is the temporal difference between two consecutive frames of a video. The higher the difference, higher will be the motion complexity of the video. This feature helps in estimating the motion content in a video sequence. It is expected that videos with higher frame rates have higher motion complexity and vice versa. The simplest way to estimate temporal frame difference is by taking the difference between pixel intensities of two consecutive frames, where pixel intensities lie within the range of 0-255. The average, maxima and minima of these temporal differences can be manipulated further for more accurate quality estimation.  
\item Temporal difference measurements along with the frame rate, the duration of the video sequence under estimation and the total number of frames should be included in overall quality estimation of a compressed video sequence.  
\item Furthermore, temporal difference helps in highlighting the scene changes in a video sequence. For example, if there is a scene change in a video sequence then the temporal difference between the consecutive frames where the scene change happens will be much higher compared to rest of the video sequence. Scene changes in compressed videos can lead to a different perceptual quality and if highlighted then they can help in better estimation of video quality.  
\end{itemize}
The aforementioned measures can be part of any FR-VQM to enhance the overall quality estimation and it will help in a better correlation between the VQMs' measurements and the DMOS.
\section{Conclusion}
This paper presented the impact of HFR on the perceptual quality of compressed FHD videos. The FHD video content at three frame rates, obtained from the BVI-HFR database, was compressed using H.265/HEVC and VP9 encoders at five QP levels. A detailed subjective quality assessment of the compressed videos for both encoders and individual frame rates was performed, which resulted in DMOS values as subjective measurements. The benchmarking investigation of the FR-objective quality assessment using 11 state-of-the-art metrics was conducted to show the correlation between the subjective and objective models. We showed that the performance of H.265/HEVC for each frame rate at each QP level is better than VP9. With an increase in frame rate, the perceptual quality in terms of FR-VQA metrics such as PSNR, SSIM, WSNR, UQI, VIF, and VMAF, also increased for both H.265/HEVC and VP9. After performing statistical evaluation to validate both models in terms of cc, FR-VQA metrics for both H.265/HEVC and VP9 is recommended for compressed FHD contents. We also provide a recommendation for enhancing the quality estimation of full-reference (FR) video quality measurements (VQMs) is presented after the extensive investigation.  Furthermore, in case of a network with low-bit requirement, the recommended QP level for each frame rate was shown for each encoder that can reflect a better visual quality to the users.  
\par 
There are several relevant research concerns to be addressed in the future. Additional investigation on the impact of HFRs, such as 90fps and 120fps, at different QP levels for 4K and 8K UHD video contents will be significant and interesting. Besides, developing new objective quality metrics for such environments is desirable. 
\section*{Acknowledgment}

The authors would like to thank...

\ifCLASSOPTIONcaptionsoff
  \newpage
\fi



%
\bibliographystyle{IEEEtran}
\bibliography{bare_jrnl}

\begin{IEEEbiography} [{\includegraphics[width=1.3in,height=1.2in,clip,keepaspectratio]{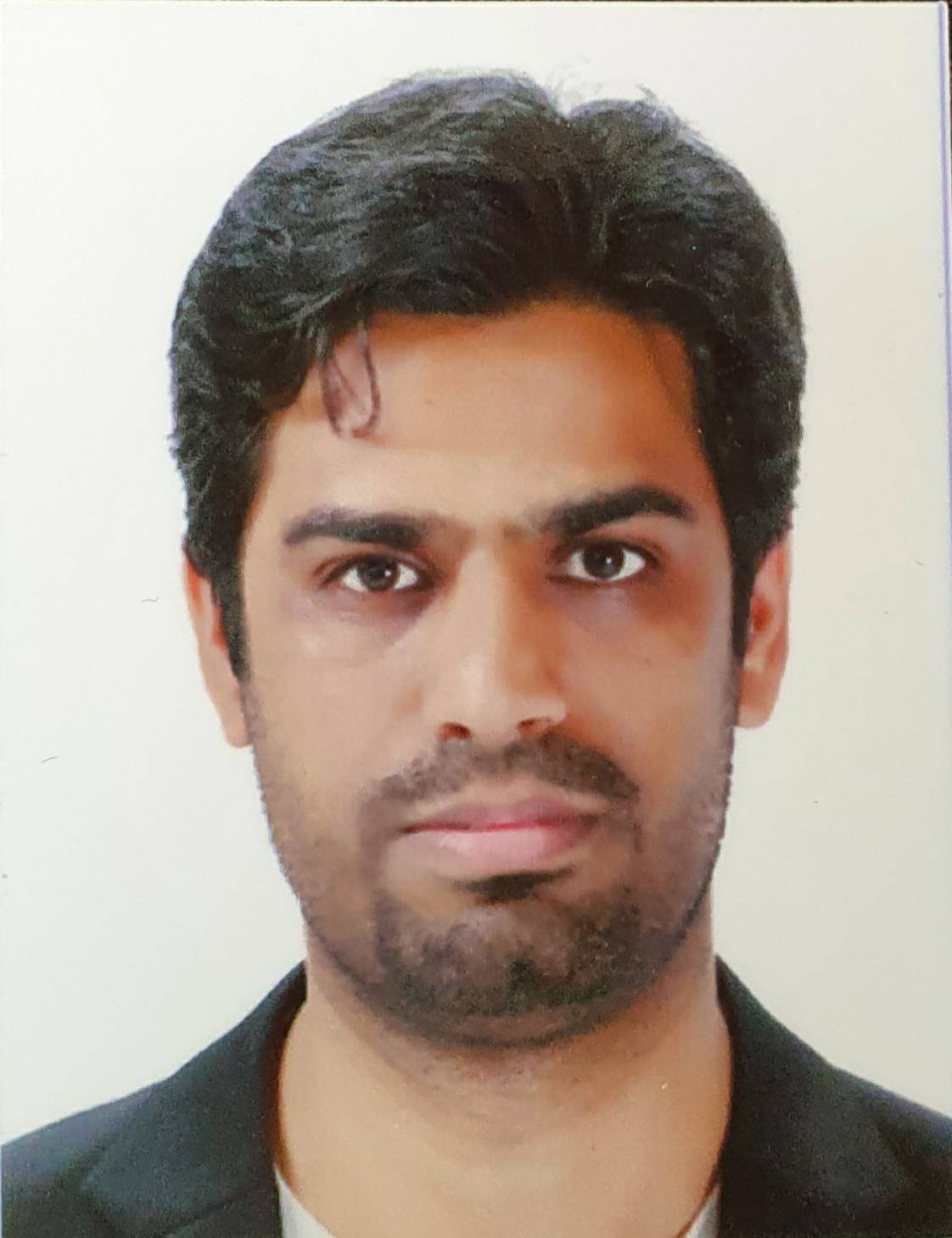}}]{Tariq Rahim}
is a Ph.D. student at Wireless and Emerging Network System Laboratory (WENS Lab.) in Department of IT Convergence Engineering, Kumoh National Institute of Technology, Republic of Korea. He has completed Master in Information and Communication Engineering from Beijing Institute of Technology, PR. China 2017. His research interests include image and video processing and quality of experience (QoE) for high resolution and high frame rate videos.
\end{IEEEbiography}

\begin{IEEEbiography} [{\includegraphics[width=1.2in,height=1.15in,clip,keepaspectratio]{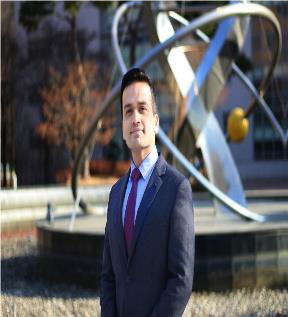}}]{Muhammad Arslan Usman}
received B.S. degree in Electrical Engineering (Telecommunications) from COMSATS University, Lahore, Pakistan, in 2010 and M.S. Degree in Electrical Engineering (Signal Processing) from Blekinge Tekniska Högskola (BTH), Karlskrona, Sweden, in 2013. I completed my PhD degree in the field of IT Convergence Engineering from Kumoh National Institute of Technology, South Korea, in Jan. 2018, and later I continued in the same research group as a postdoctoral research fellow for a period of 4 months. My research interests include Quality of Experience (QoE) estimation, modelling and management for medical and general videos, object classification and detection in computer vision applications, and next generation wireless networks including non-orthogonal multiple access. Currently I am working as a Postdoctoral research fellow with Wireless Multimedia and Networking (WMN) research group, Kingston University, United Kingdom.
\end{IEEEbiography}

\begin{IEEEbiography} [{\includegraphics[width=1in,height=1.35in,clip,keepaspectratio]{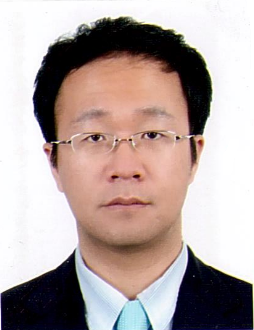}}]{Soo Young Shin}
received his BS, MS, and PhD degrees in Electrical Engineering and Computer Science from Seoul National University, Korea in 1999, 2001, and 2006, respectively. He was a visiting scholar in FUN Lab at University of Washington, US, from July 2006 to June 2007. After 3 years working in WiMAX design Lab. of Samsung Electronics, he is now associate professor in School of Electronics in Kumoh National Institute of Technology since September 2010. His research interests include wireless LAN, WPAN, WBAN, wireless mesh network, sensor networks, coexistence among wireless networks, industrial and military network, cognitive radio networks, and next generation mobile wireless broadband networks. 
\end{IEEEbiography}

\end{document}